\documentclass[iop]{emulateapj}
\usepackage[outdir=./]{epstopdf}
\usepackage{graphicx}
\usepackage{threeparttable}
\usepackage{rotating}
\usepackage{afterpage}
\usepackage[load-configurations=astronomy]{siunitx}
\DeclareSIUnit\micron{\ensuremath{\mathrm{\micro\meter}}}
\DeclareSIUnit\sig{\ensuremath{\mathrm{\sigma}}}
\DeclareSIUnit\mag{mag}
\DeclareSIUnit\arcsec{\ensuremath{\mathrm{^{\prime\prime}}}}
\DeclareSIUnit\parsec{pc}
\DeclareSIUnit\jansky{mJy}
\DeclareSIUnit\deg{deg}
\DeclareSIUnit\yr{yr}
\DeclareSIUnit\Lsun{\ensuremath{\mathrm{L_\odot}}}
\DeclareSIUnit\Msun{\ensuremath{\mathrm{M_\odot}}}
\usepackage{natbib}
\shorttitle{``Red" But Not ``Dead"}
\shortauthors{Runge, Yan}

\begin{document}

\title{``Red" but Not ``Dead": Actively Star-forming Brightest Cluster Galaxies at Low Redshifts}

\author{James Runge\footnotemark[\textdagger]}
\footnotetext[\textdagger]{email: jmr24f@mail.missouri.edu}
\author{Haojing Yan\footnotemark[$\ddagger$]}
\footnotetext[$\ddagger$]{email: yanha@missouri.edu}
\affil{Department of Physics and Astronomy, University of Missouri - Columbia}

\begin{abstract}
Brightest Cluster Galaxies (BCGs) are believed to have assembled most of their stars early in time and, therefore, should be passively evolving at low redshifts and appear ``red-and-dead." However, there have been reports that a minority of low-redshift BCGs still have ongoing star formation rates (SFR) of a few to even $\sim$100 $M_\odot/yr$. Such BCGs are found in ``cool-core" (``CC") clusters, and their star formation is thought to be fueled by ``cooling flow." To further investigate the implications of low-redshift, star-forming BCGs, we perform a systematic search using the 22$\mu$m data (``W4" band) from the Wide-field Infrared Survey Explorer (WISE) on the GMBCG catalog, which contains 55,424 BCGs at $0.1\lesssim z\lesssim 0.55$ identified in the Sloan Digital Sky Survey (SDSS). Our sample consists of 389 BCGs that are bright in W4 (``W4BCGs"), most being brighter than 5 mJy. While some ($\lesssim 20\%$) might host AGN, most W4BCGs should owe their strong mid-IR emissions to dust-enshrouded star formation. Their median total IR luminosity ($L_{IR}$) is $5\times10^{11} L_{\odot}$ (SFR $\sim$50 $M_{\odot}/yr)$, and 27\% of the whole sample has $L_{IR}>10^{12} L_{\odot}$ (SFR $>$100 $M_{\odot}/yr$). Using ten W4BCGs that have {\it Chandra} X-ray data, we show that seven of them are possibly in CC clusters. However, in most cases (five out of seven) the mass deposition rate cannot account for the observed SFR. This casts doubt to the idea that cooling flows are the cause of the star formation in non-quiescent BCGs.	
  
\end{abstract}


\section{Introduction}
A Brightest Cluster Galaxy (BCG), as the name implies, resides within a galaxy cluster and is its brightest member. BCGs are among the most luminous and the most massive galaxies in the low-redshift universe, usually have little ongoing star formation, and are dominated by old stellar populations (e.g., \citealt{Dub98}). It is believed that they have assembled most of their stellar masses before $z\sim 3$ (e.g., \citealt{DeL06}), and have been passively evolving ever since. For this reason, they are among the so-called ``red-and-dead" galaxy population.

BCGs being largely quiescent in the low-redshift universe is consistent with the general picture of ``downsizing" galaxy evolution (\citealt{Cowie96}), where the bulk of the star formation activities in the universe shift from high mass galaxies to low mass ones as the  universe evolves. On the other hand, it has been known for over a decade that some BCGs at low redshifts still exhibit significant star formation. Such BCGs are in ``cool-core" clusters, whose intra-cluster medium (ICM) has a  temperature gradient such that materials can be funneled to the central region where the BCGs reside and presumably can fuel the observed star formation (e.g., \citealt{ODea05,Vik07,Santos08,Fog15}; see also \citealt{Don15} for recent discussions). However, it is unclear what fraction of BCGs still have ongoing star formation.

In this paper, we present a systematic study of star-forming BCGs, using the data from the Sloan Digital Sky Survey (SDSS; \citealt{York2000}) and the Wide-field Infrared Survey Explorer (WISE; \citealt{Wright2010}). We make use of the GMBCG catalog of \cite{Hao2010}, which is the largest BCG catalog available to date, and select those BCGs that potentially have strong ongoing star formation activities based on their properties in the mid-IR bands of WISE. Our goal is to shed new light to the understanding of star-forming BCGs as a whole: how rare they are, how high their star formation rates (SFRs) can be, whether they have different properties in other aspects as compared to  the vast majority of BCGs that are quiescent, and whether residing in ``cool-core" clusters is a satisfactory explanation to their SFRs.

\begin{figure*}[th]
	
	\raggedright
	\noindent\hspace{0.0in}{\Large \textbf{W1}}
	\noindent\hspace{1.1in}{\Large \textbf{W2}}
	\noindent\hspace{1.1in}{\Large \textbf{W3}}
	\noindent\hspace{1.1in}{\Large \textbf{W4}}
	\vspace{4pt}
	\plotone{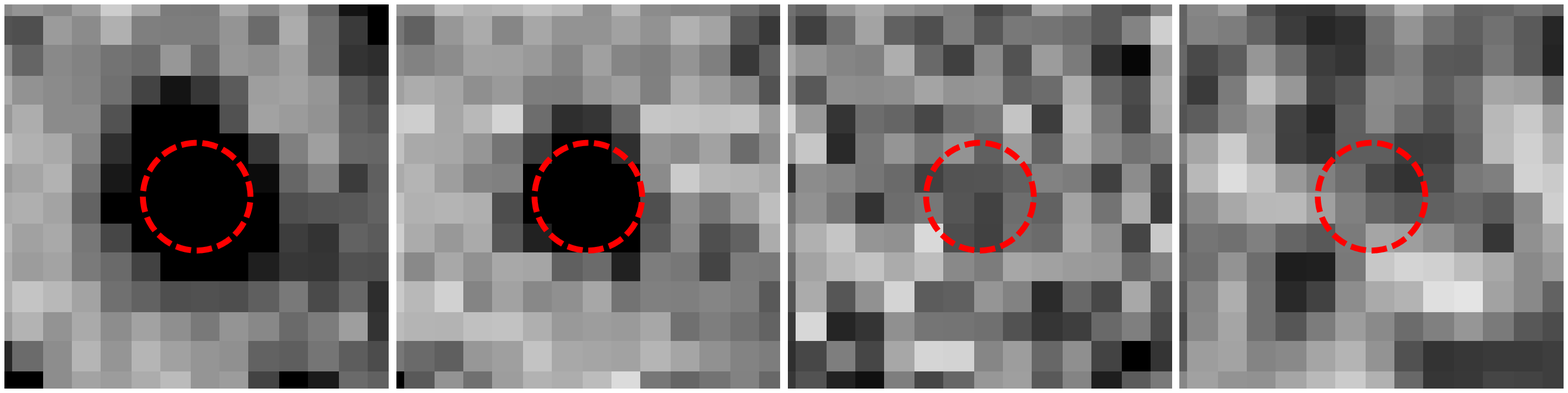}
	
	\vspace{8pt}
	
	\plotone{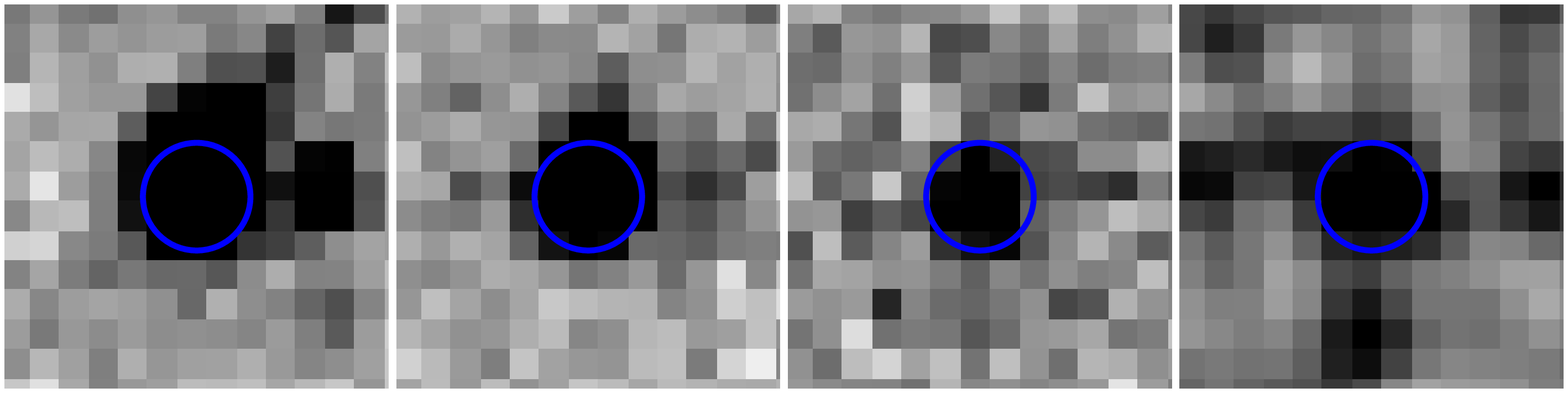}
	
	\vspace{2pt}
	
	\caption{Examples showing WISE 4-band images of a BCG that is undetected in the W4 band (top) and one that is detected (bottom). These image cutouts are made from the unWISE products. The circles are centered on the reported SDSS positions, and are 10\arcsec ~in radius.}
	\label{fig:det}
\end{figure*}
This paper is organized as follows. In Section 2, we briefly describe the data that we use to select star-forming BCGs. The details of the selection process is given in Section 3, and these are followed by Section 4 where we show various diagnostics to separate AGN activity from star formation. In Section 5, we analyze a small subsample of star-forming BCGs that have archival X-ray data that allow us to address various questions regarding their connection to cool-core clusters. We present a discussion of our findings in Section 6, and a summary in Section 7.

Throughout this paper, we adopt the concordant $\Lambda$CDM cosmological model of $H_0=70$~ Mpc$^{-1}$km~s$^{-1}$, $\Omega_M=0.27$, and $\Omega_{\Lambda}=0.73$. All magnitudes are in AB system unless otherwise noted.

\section{Data}
The critical data sets used to select star-forming BCGs are the SDSS-based GMBCG catalog and the WISE all-sky survey data. In particular, we adopt the ``un-blurred" version of the WISE data (also known as ``unWISE") of \cite{Lang12014} for this study. A small number of such selected BCGs also have far-IR (FIR) data from {\it Herschel}\, or X-ray data from {\it Chandra}, which we used for further analysis. All these data are briefly described below.
\subsection{GMBCG Catalog}
The GMBCG Catalog (\citealt{Hao2010}) consists of 55,424 rich galaxy clusters found
by using the Gaussian Mixture Brightest Cluster Galaxy (GMBCG) algorithm on the SDSS data in the seventh data release (DR7). This algorithm detects clusters by identifying the BCG and the red sequence galaxies in its vicinity and calculating the clustering strength, a measure of the surface density of cluster galaxies at the BCG position. \cite{Hao2010} apply this method to the Legacy Survey Area of SDSS DR7, which covers 7,300~deg$^2$ of the North Galactic Cap and 740~deg$^2$ from three stripes in the South Galactic Cap, and obtain their cluster catalog across the redshift range $0.1\lesssim z \lesssim 0.55$.

This GMBCG catalog contains the positions of the identified BCGs along with their redshifts and photometry. The redshifts are either spectroscopic redshifts ($\sim$20,000 objects) or photometric redshifts (see \citealt{Hao2010} for details).

\subsection{WISE and unWISE}
The nominal WISE mission mapped the entire sky in 2010 in four near-to-mid-IR bands, namely, W1, W2, W3, and W4, whose central wavelengths are 3.4, 4.6, 12, and 22~$\mu$m, respectively. The spatial resolutions in these four bands are
\SI{6.1}{\arcsec}, \SI{6.4}{\arcsec}, \SI{6.5}{\arcsec}, and \SI{12.0}{\arcsec}, respectively.
The nominal 5$\sigma$ limits in these bands are 0.068, 0.098, 0.86 and \SI{5.4}{\jansky}, respectively (\citealt{Wright2010}).

The offically released images of WISE (``AllWISE") were intentionally convolved by the point spread functions (PSFs) during the co-adding process. While this process is appropriate for isolated point sources, it reduces the resolution of the images and thus exacerbates the blending problem. To remedy this problem, unWISE\footnote[1]{http://unwise.me} (\citealt{Lang12014})
\textquotedblleft un-blurs" these images to produce the final stacks that preserve the native spatial resolutions.

Along with the un-blurred images, unWISE also provides a catalog of WISE photometry based on ``forced photometry" using $\sim$ 400 million SDSS DR10 objects as the morphological templates to fit the WISE source light profiles (\citealt{Lang22014}). Since the GMBCG catalog is based upon the same SDSS data (albeit in an earlier data release), all of our objects appear in the unWISE forced photometry catalog. Therefore, we adopted the unWISE images for visual verification and its forced photometry for quantitative analysis.
\begin{figure*}

	\plottwo{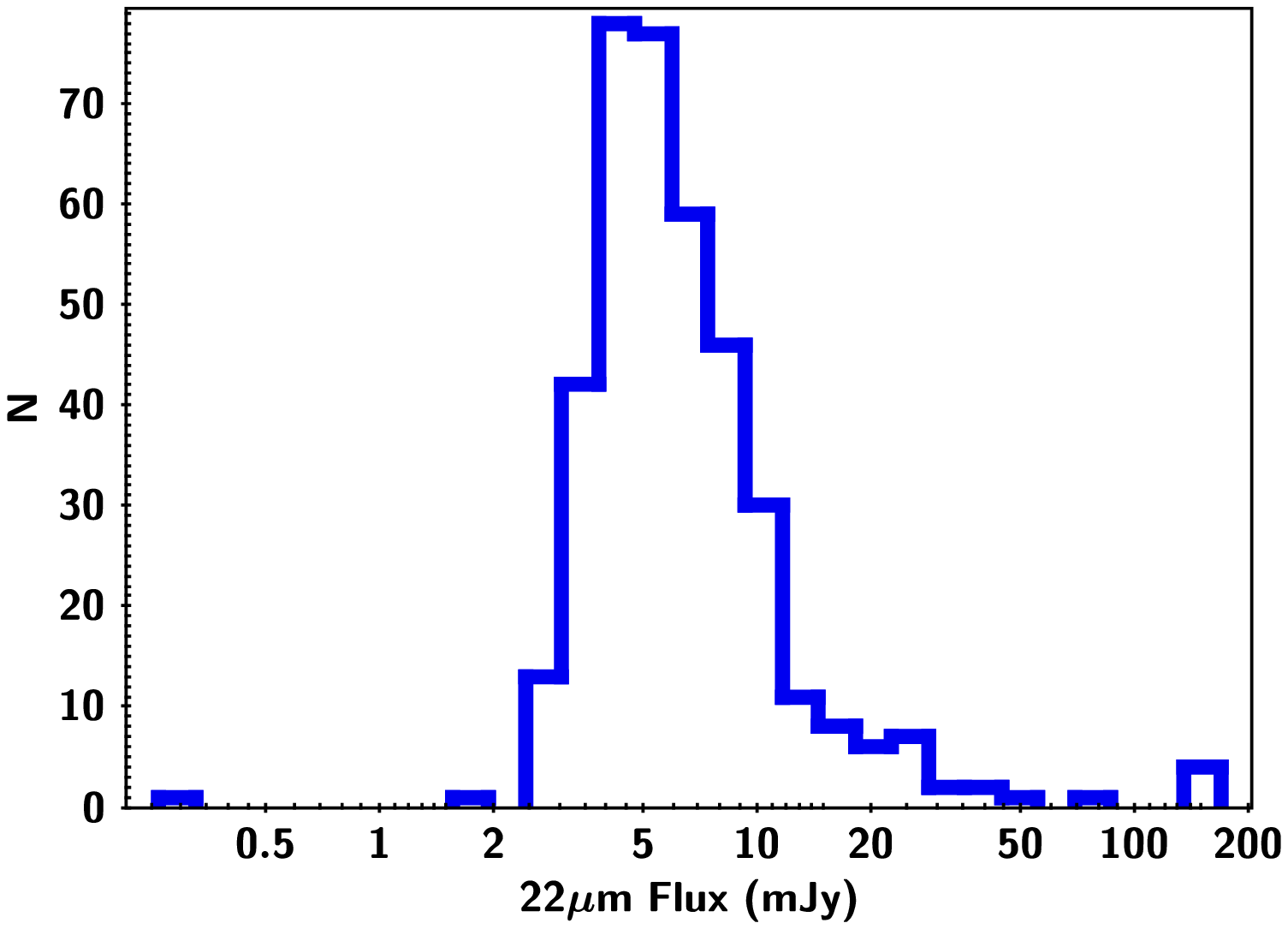}{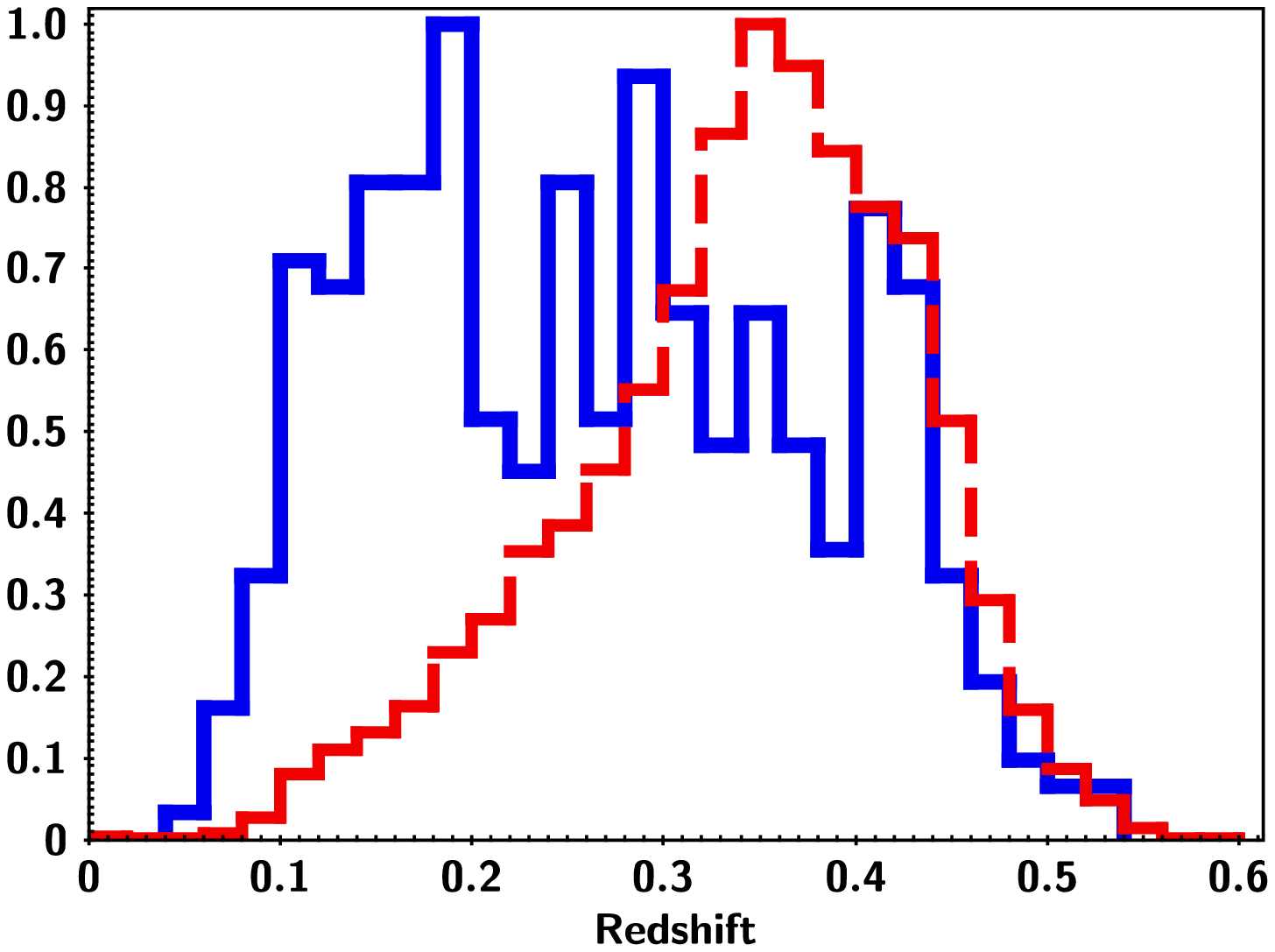}
	\caption{Left: W4 (\SI{22}{\micron}) flux density distribution of the W4-detected BCGs (W4BCGs). Right: Normalized redshift distribution for the entire GMBCG catalog (red dashed line) and the subset of W4BCGs (blue solid line).}
	\label{fig:histos}
\end{figure*}

\subsection{Herschel data}
In order to further study the star formation properties of the selected BCGs, we also used the archival FIR data taken by the Spectral and Photometric Imaging Receiver (SPIRE; \citealt{Griffin10}) on {\it Herschel}\, Space Observatory (\citealt{Pilbratt10}). While only a small number of objects have these SPIRE data, they offer a valuable reference that we will detail in \S 4.3.

 
 Specifically, we made use of the SPIRE three-band (250, 350 and 500~$\mu$m) photometry from the following {\it Herschel}\, very wide-field surveys whose SPIRE data are now publicly available, namely, the \textit{Herschel} Multi-tiered  Extragalactic  Survey  (HerMES;  \citealt{Oliver12,Roseboom10,Smith12, Viero13,Wang14}), the \textit{Herschel} Stripe 82 Survey (HerS; \citealt{Viero14}) and the {\it Herschel}  Astrophysical Terahertz Large Area Survey  (H-ATLAS; \citealt{Eales10b,Val16}). Both HerMES and H-ATLAS have catalogs available that include flux measurements. For the HerS data, we measured the source flux on the SPIRE images using HIPE (\citealt{Ott10}) following the procedure for source extraction and photometry outlined in the SPIRE data manual\footnote[2]{http://herschel.esac.esa.int/hcss-doc-12.0/}.   In total, these surveys cover ~340 deg$^2$.

\subsection{Chandra Data}
In order to investigate possible cool-core properties of our sample, X-ray data is necessary. Therefore, we used the public \textit{Chandra} X-ray data provided by the Chandra Data Archive. Both Primary and Secondary products were retrieved for each available observation. These data provide a spatial resolution of \SI{0.5}{\arcsecond} and cover an energy range of 0.1-10 keV.

Standard data processing was carried out starting from the level 1 event files using \texttt{CIAO} 4.8.2 (\citealt{Fruscione06}) with CALDB 4.7.0 of the \textit{Chandra} Calibration Database. The reprocessing script \textit{chandra\_repro} was used to reprocess the data and create level 2 event files. When observations were taken in the VFAINT mode, the parameter \textit{check\_vf\_pha} was set to ``yes'' in order to remove background events likely caused by cosmic rays. Background estimates were taken in the same field away from the central X-ray peak and clear of any other X-ray sources.

For X-ray spectra, we followed a procedure similar to that of \citealt{Mol16} (hereafter Mol16). The X-ray spectra for the BCG were processed from the level 2 event files using \textit{specextract} in \texttt{CIAO}. A 40 kpc region centered on the BCG was chosen for the extraction. There could be a complication in this analysis if the BCG is an X-ray AGN, in which case the X-ray spectrum might be dominated by the AGN rather than the heated ICM. To solve this potential problem, we performed a separate analysis by following Mol16 and masking the central region. Unlike in Mol16 where a circular region of 2\arcsec\, in radius is masked, we chose to only mask out the inner 2~kpc of the BCG, as choosing a global value of 2\arcsec\, would result in masking out the bulk of X-ray flux for some of our sources. Background spectra were also processed at up to three different regions away from the X-ray peak and any other X-ray sources.

\section{Sample of BCGs with strong mid-IR emission}
We searched for BCGs with ongoing star formation by identifying those that have secure mid-IR detections in the WISE W4-band at 22~$\mu$m. This will reveal dust-embedded star formation in BCGs, and thus is complementary to the method that aims at identifying unobscured star formation through UV emissions, such as some of those reported by \cite{Don15}. Here we describe our sample in detail.
\subsection{Initial Selection}

To construct a catalog of possible sources that have W4 detection, the GMBCG catalog was cross-matched with the unWISE forced photometry catalog using a matching radius of \SI{5}{\arcsecond}, which is slightly less than half the spatial resolution of W4. We further required that a matched object should have $S/N\geq5$ in W4 as reported in the unWISE SDSS forced photometry catalog. This resulted in 1,323 BCGs in our initial sample.

To ensure the sample robustness, we visually inspected the images of all these initial candidates. We found that a large number of the reported W4-detections were actually false-positives due to various reasons, such as image defects, noise spikes, artifacts produced by a bright neighbor, etc. After rejecting these contaminators, 458 BCGs survived. As an example, Fig.~\ref{fig:det} shows the WISE image stamps of one that is not detected in W4 and one with a real W4-detection. 
\subsection{Sample Verification}
Obviously, W4-detected BCGs are only a small fraction of the entire GMBCG sample. Therefore, we must consider possible contamination to the GMBCG sample, or in other words, whether these W4-detected objects derived from the GMBCG sample are BCGs at all. To address this question, we further verified the legitimacy of these 458 candidate objects on a one-by-one basis. This verification was to decide whether a candidate is in a cluster environment, and if yes, whether it is the BCG of the cluster. Our intention was not to invent a new cluster finding algorithm, but to perform an independent ``sanity check'' on the claimed BCGs.

The verification consisted of two steps. First, we worked under the assumption that the photometric redshifts that the GMBCG catalog relies on are accurate enough for its purpose. We used the SDSS DR7 data, the same as what the GMBCG catalog is based. For each candidate BCG, we retrieved the objects within a 3\arcmin\, radius around it, and retained only those whose photometric redshifts (as reported in the SDSS DR7) were within $\pm 0.02$ of the redshift of the candidate BCG (as quoted in the GMBCG catalog and is the same as in the SDSS DR7). The retained objects were considered as the members of the candidate cluster. This redshift range was adopted because it is the reported accuracy (1$\sigma$) of the SDSS DR7 photometric redshifts. We then constructed the $i$ vs. $(g-r)$ color-magnitude diagram, and checked whether we could see a ``ridge line'' indicative of a red sequence. If a red sequence was seen, we checked whether the current candidate BCG was the correct identification of BCG, i.e., whether it was the brightest one (in $i$-band) among all members.

After this step, we confirmed that 383 objects survived. Among the 75 dubious cases,  four could hardly be called clusters because they only had a few members ($<8$), and thus must be rejected. These four sources were at the high-redshift end of the catalog. One other case was a misidentification, and actually must be part of Abell 1689 (whose BCG is already in the W4BCG sample) and thus must also be removed. The other 70 objects were in clusters with a clear red sequence, however they were in fact not the BCGs. Therefore, we identified the ``new" BCG for each of these 70 cases by finding the brightest member, and conducted all the previous steps reported above on these ``new" BCGs. Of all these 70 objects, only six have reliable W4 detections. We included these six objects into our sample, and
thus our final sample consists of 389 objects in total.

\subsection{General Properties and Subdivision of the Final Sample}
   
~Fig.~\ref{fig:histos} shows the distributions of their W4 flux densities and redshifts. At these redshifts, the W4 emissions are still in the rest-frame mid-IR, and  must be originated from heated dust instead of stellar continuum. These W4-detected BCGs (hereafter ``W4BCGs") comprise a special population at odd with the general picture of BCGs that they are old, passively evolving galaxies. Therefore, we aim to understand the nature of these exceptions.

In order to investigate whether the occurrence of W4BCG could be dependent of the cluster richness, we divide our final W4BCG sample into two categories based on the reported GMBCG cluster richness (``$N^{scaled}_{gals}$'')  in \cite{Hao2010}. We adopt $N^{scaled}_{gals}=15$ as the criterion, and refer to those clusters with $N^{scaled}_{gals}\geq 15$ as ``rich'' clusters and those with $N^{scaled}_{gals}<15$ as ``poor'' clusters. The rich clusters are 28\% of the entire GMBCG sample, while the poor clusters make of 72\%. The corresponding W4BCGs are subsequently divided into the W4BCG-R (108 objects, or 27.8\% of the total 389 W4BCGs) and the W4BCG-P (281 objects, or 72.2\%) subsamples, respectively.

\section{Data Analysis}
There are two possible causes to the 
heated dust emissions of these W4BCGs in the mid-IR, namely, active ongoing star formation or AGN activities. In this section, we investigate which of these two mechanisms is the more probable cause.
\subsection{Possible AGN Hosts}

To understand whether any of our W4BCGs could possibly host an AGN, we performed two diagnostics, which are based on the WISE color selection and the BPT diagram method, respectively. We note that being diagnosed as an AGN host by either method does not necessarily mean that the mid-IR emission in W4 must be dominated by AGN heating. However, if we do not find any AGN activity by either method, it is very plausible that the mid-IR emission is mainly driven by the heating of star formation.

\subsubsection{WISE Color Diagnostics}

\begin{figure}[t]
		
		\centerline{\includegraphics[width=250pt,trim=5pt 0 0 0,clip]{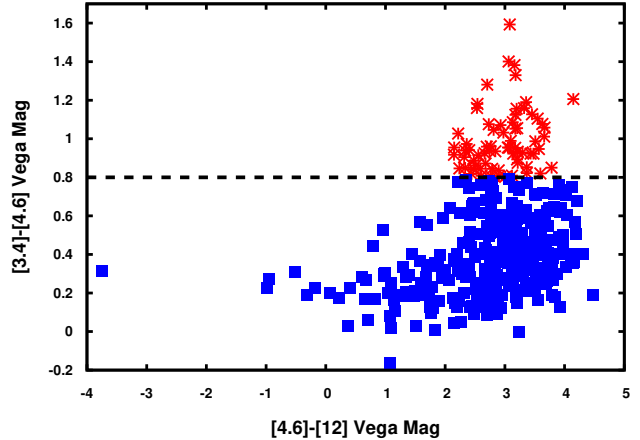}}
		
	\caption{WISE color diagnostics of W4BCGs. The dashed line at $W1-W2=0.8$~mag (\citealt{Stern2012}) separates AGN (red asterisks) and non-AGN (blue squares). Only 69 out of the total 389 W4BCGs are possible AGN hosts by this selection. \label{fig:cc}}
\end{figure}

Using a W1-W2 versus W2-W3 WISE color-color plot has been shown to be an effective method to identify AGN (\citealt{Jarrett2011,Mateos2012,Stern2012,Assef2013}). Furthermore, it has been demonstrated that a single color criterion of $W1-W2\geq 0.8$~mag (in Vega system) provides a robust selection of AGN  (\citealt{Stern2012,Assef2013}).  We adopted this latter method in our analysis, and the result is shown in Fig~\ref{fig:cc}.~We find that only 69 out of the total 389 W4BCGs and 12 of the 108 W4BCG-Rs satisfy this criterion (17.7\% and 11.1\% respectively), or in other words, most W4BCGs should be dominated by starbursts.

\subsubsection{BPT Diagram}

\begin{figure}[t]
	\centerline{\includegraphics[width=250pt]{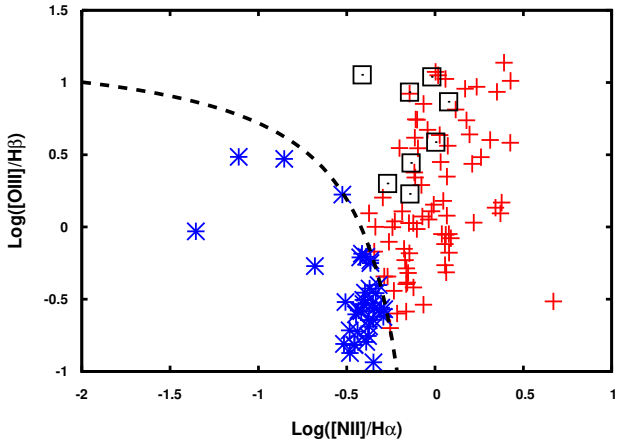}}
	\caption{BPT diagnostics of W4BCGs. The dashed curve represents the criterion of Kauffman et al. (2003) that separates AGN-dominated objects (red pluses) and star-forming-dominated objects (blue asterisks). The black open squares (8 in total) indicate those that are also deemed to be AGN hosts based on the WISE color selection (see Fig.~\ref{fig:cc}.)}\label{fig:BPT}
\end{figure}

BPT diagrams are a set of diagnostic diagrams using emission lines to determine the ionization mechanism of nebular gas. The most commonly used diagram is [OIII]5007/H$\beta$ versus [NII]6584/H$\alpha$ (\citealt*{Baldwin1981}), which is what we used in our analysis. Various dividing curves have been proposed to separate AGN from star-forming galaxies (e.g. \citealt{Kewley01,Kauffmann03}). The curve of \cite{Kauffmann03}, which is shown in Eq.~\ref{eq:K} below, is the most aggressive in selecting AGN, and hence we adopted this selection criterion in order to be conservative in attributing W4BCGs to starburst:

\begin{equation} 
\label{eq:K}
log([OIII]/H\beta) = 0.61/(log([NII]/H\alpha)- 0.05) + 1.3. 
\end{equation}

For emission line measurements, we used the MPA-JHU(Max-Planck
Institute for Astrophysics - John Hopkins University) ``value-added" DR7 catalog of spectrum measurements (\citealt{Brinchmann04,Kauff03,Tremonti04}) based on the SDSS DR7 data. A cross-match to the MPA-JHU catalog (within a radius of \SI{5}{\arcsec} of the GMBCG reported position) resulted in 123 objects with emission line measurements for  all four lines needed for the BPT diagnostic, which is shown in Fig~\ref{fig:BPT}. Within this subsample of 118 W4BCGs, 84 are deemed to host AGN (29 are W4BCG-R).

\subsection{Morphology}
\begin{figure}[t]
	\plotone{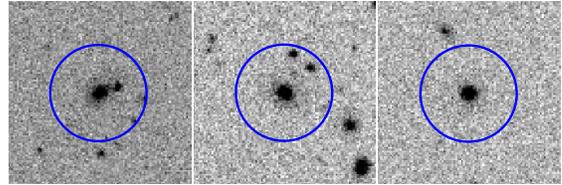}
	\caption{SDSS \textit{i}\arcmin~band images illustrating the different morphology subsets. Left: A BCG with obvious merger properties. Middle: A BCG with a neighbor within 10\arcsec ~but no obvious merger properties. Right: A BCG with no other galaxies within a 10\arcsec ~radius.}\label{fig:MNS}
\end{figure}

The morphologies of the W4BCGs may provide additional information to reveal the nature of their mid-IR emissions. In particular, we are interested in understanding whether merger could be relevant, regardless of the exact heating source being AGN activities or star formation. For this purpose, the \texttt{SDSS} \textit{i}\arcmin~band images were examined. The W4BCGs were then divided into three different categories: ``Merger", where a recent or ongoing merger is evident as shown by disturbed morphology; ``Close Neighbor", where there is no clear sign of merger but there is at least one galaxy within 10\arcsec~(even though this could be due to projection by chance); and ``Single", where there is no sign of merger and no other galaxies within 10\arcsec~to the SDSS depth. Some examples are shown in Fig.~\ref{fig:MNS}. We note that 10\arcsec~corresponds to 18\textendash60 kpc at the redshifts of the W4BCGs.

The statistics are listed in Table~\ref{tab:morph}, which shows that the majority of W4BCGs do not exhibit obvious merger properties. Therefore, it is safe to conclude that the mid-IR emission of a W4BCG is independent of whether it is interacting with others.

\begin{table}[t]
	\centering
	\caption{Morphology of W4 detected BCGs based on SDSS \textit{i'} images}
	\begin{tabular}{l c}
		
		Type & Count \\
		\hline
		\vspace{-0.6em}\\
		Merger & 69 \\
		Close Neighbor & 145  \\
		Single & 175 \\
		
	\end{tabular}\label{tab:morph}
\end{table}\label{tab:m}
\subsection{Star Formation Rates}
 \begin{figure*}[t]
 	\centerline{\includegraphics[height=350pt,width=500pt,trim=5pt 40pt 25pt 0,clip]{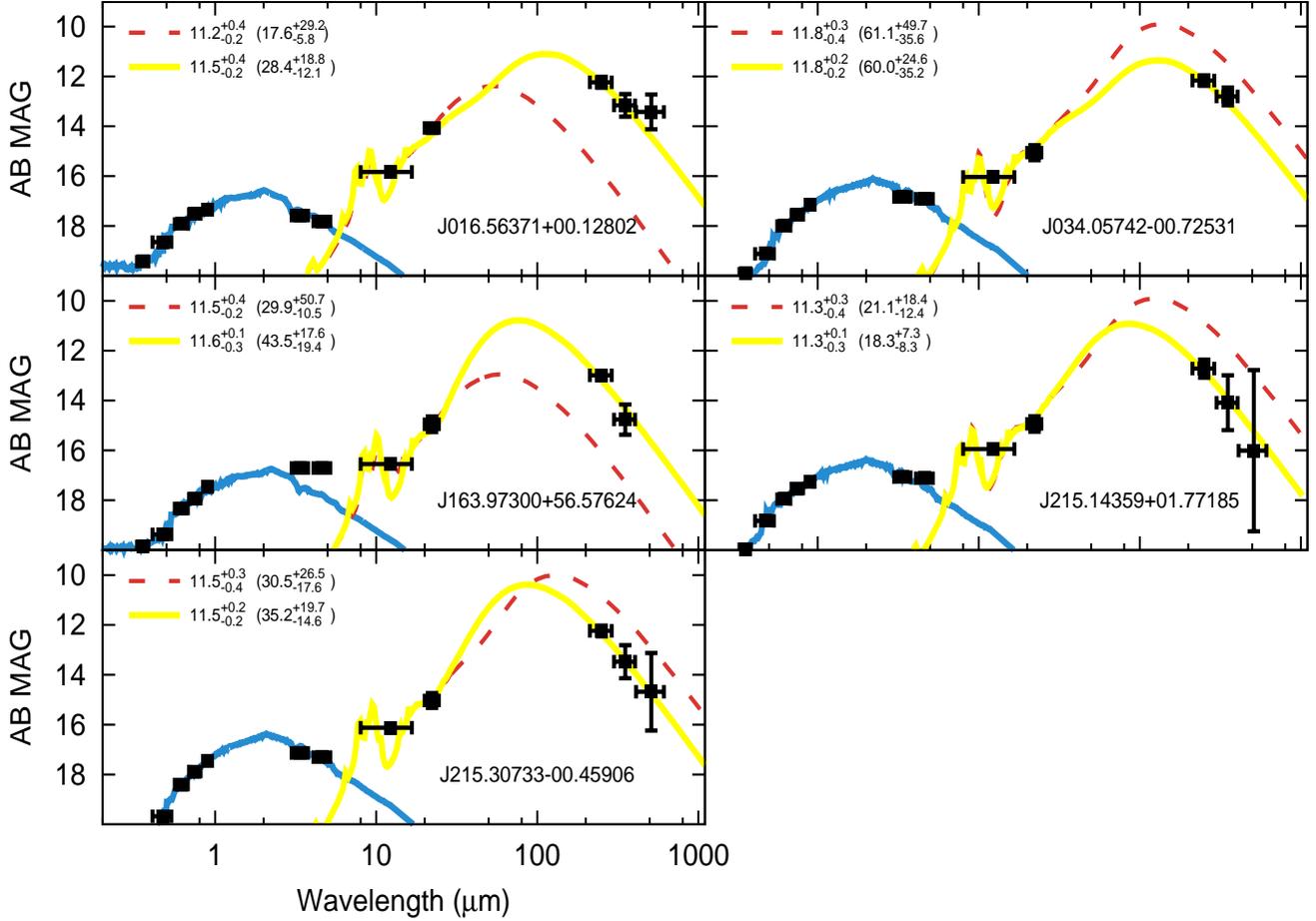}}
 	\caption{Spectral energy distributions of the subsample of five W4BCGs that have {\it Herschel}\, data. The fitting of their stellar populations (shown by the blue curve) is the same as in the other W4BCGs, which incorporates both the SDSS photometry in optical and W1 and W2 in near-IR (see text for details). Two fitting schemes in the mid-to-far-IR regime are shown: one only using WISE data (dashed red) and one including \textit{Herschel} data (solid yellow). The blue curve shows the SED fitting results of the stellar populations, which incorporate the SDSS photometry in optical and the WISE near-IR photometry in W1 and W2. The log($L_{IR}$) value is displayed in the top left for each fitting along with the corresponding SFR (in $M_\odot$/yr) in parenthesis. \label{fig:multi}}
 \end{figure*}
\begin{table*}[t!]
	\centering
	\begin{threeparttable}
		\centering
		\caption{Summary of the Subset with \textit{Herschel} Data}
		\begin{tabular}
			{lccccccc}
			\hline
			\hline
			\vspace{-0.6em}\\
			& 250$\mu$m & 350$\mu$m&500$\mu$m& Log($L_{IR}/L_{\odot}$)&$SFR_{NoSPIRE}$&Log($L_{IR}/L_{\odot}$)&$SFR_{SPIRE}$  \\
			GMBCG Catalog Name  & (mJy) & (mJy)&(mJy) & No SPIRE&($M_\odot$/yr)&w/ SPIRE&($M_\odot$/yr)  \\
			\hline
			\vspace{-0.6em}\\
			W4BCG-R &&&&&&&\\
			J034.05742-00.72531 & 49.57$\pm$5.92 & 27.54$\pm$5.40 & --& 11.79$^{+0.26}_{-0.38}$ & 61.12$^{+49.67}_{-35.60}$  & 11.78$^{+0.20}_{-0.23}$ & 59.99$^{+35.22}_{-24.56}$  \\
			\\
			W4BCG-P &&&&&&&\\
			J016.56371+00.12802 & 46.39$\pm$5.84 & 19.81$\pm$4.81  &15.50$\pm$5.42 & 11.25$^{+0.43}_{-0.17}$ & 17.61$^{+29.24}_{-5.82}$  & 11.45$^{+0.22}_{-0.24}$ & 28.41$^{+18.83}_{-12.15}$  \\
			J163.97300+56.57624 & 23.21$\pm$1.92 & 4.52$\pm$1.41  &-- & 11.48$^{+0.43}_{-0.19}$ & 29.88$^{+50.69}_{-10.52}$  & 11.64$^{+0.15}_{-0.26}$ & 43.51$^{+17.55}_{-19.43}$  \\
			J215.14359+01.77185 & 29.79$\pm$5.93 & 8.43$\pm$4.15  & 1.43$\pm$1.23                                                                                                                                    & 11.32$^{+0.27}_{-0.39}$ & 21.13$^{+18.37}_{-12.44}$  & 11.26$^{+0.15}_{-0.26}$ & 18.30$^{+7.27}_{-8.32}$    \\
			J215.30733-00.45906 & 46.00$\pm$5.64 & 14.76$\pm$4.95 & 4.87$\pm$3.00                                                                                                                                    & 11.48$^{+0.27}_{-0.37}$ & 30.52$^{+26.51}_{-17.59}$  & 11.55$^{+0.19}_{-0.23}$ & 35.19$^{+19.68}_{-14.56}$  \\
			
			\hline
		\end{tabular}\label{tab:hersc}
		\begin{tablenotes}
			\item	Note: {\it Herschel}\ SPIRE photometry are either adopted from the public data releases from the relevant teams when available (DR2 of HerMES and DR1 of H-ATLAS) or based on our own source extraction (for those objects in HerS) using HIPE (\citealt{Ott10}).
		\end{tablenotes}
	\end{threeparttable}
\end{table*}\label{tab:hers}
The analysis above clearly shows that most W4BCGs are not AGN hosts, and hence their mid-IR emission can only come from dust heated by strong star formation.  Again, this is contrary to the general picture that BCGs are ``red-and-dead" galaxies that have ceased their star formation long ago.  This also leads to the question whether the mid-IR emissions of those AGN-hosting W4BCGs are due to AGN heating at all, as our data currently available cannot provide an unambiguous answer. In this section, we attempt to derive the star formation rates of the W4BCGs as a whole, assuming that the AGN contribution to their mid-IR emissions is negligible. In our later discussion, we examine whether this assumption is reasonable.

 \begin{figure*}[th]
 	
 	\centerline{\includegraphics[width=500pt,trim=0 15pt 0 0,clip]{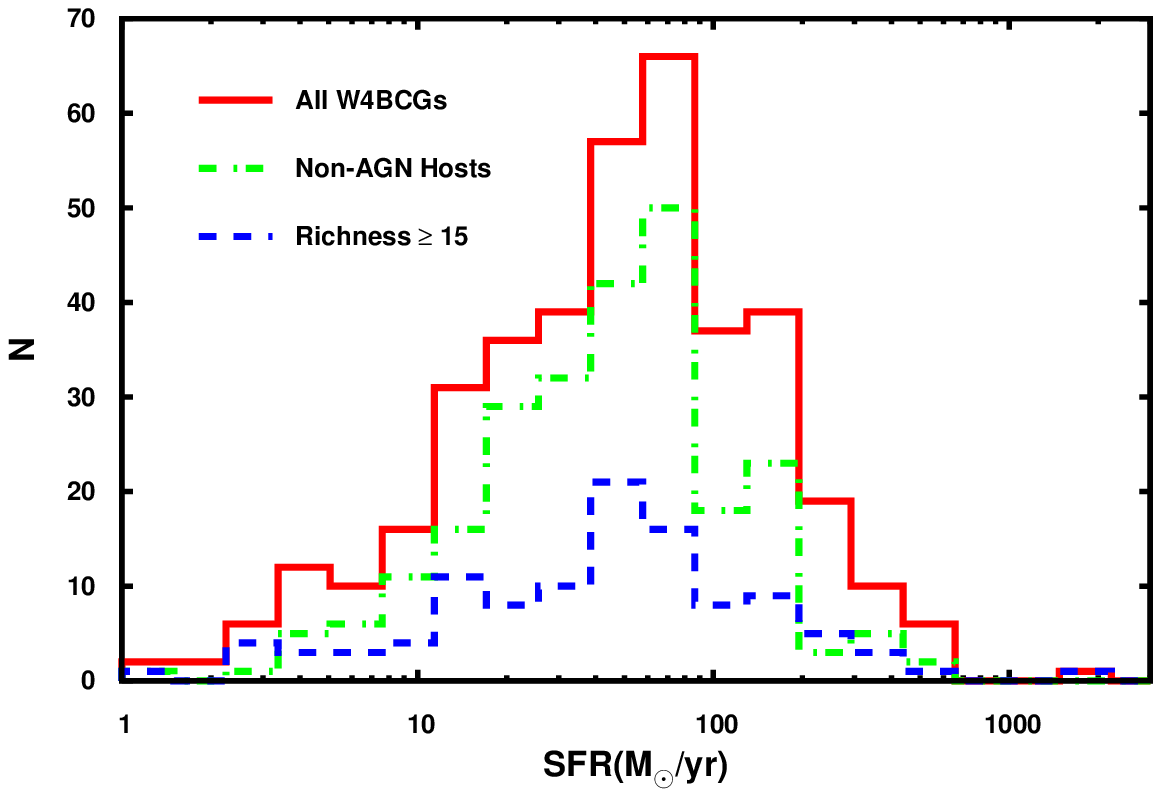}}
 	
 	\caption{A histogram of the derived SFR for the entire W4BCGs sample (red solid line), the non-AGN host subset (green dot-dashed), and the subset with richness $\geq$15 (blue dashed line; regardless of hosting AGN or not).  \label{fig:sfrhisto}}
 \end{figure*}
\subsubsection{SED Fitting}
A common method to derive IR-based SFR is to calculate the total IR luminosity ($L_{IR}$) over the conventional range of 8--1000~$\mu$m and then to infer the SFR by using the relation of \cite{Kennicut98} as follows:

\begin{equation}\label{eq:Ken}
SFR[M_\odot/yr] = 1.0\times10^{-10} L_{IR}[L_\odot].
\end{equation}
We note that the coefficient in the above equation is after adjusting to a \cite{Chab03} initial mass function (IMF) and the derived SFR is a factor of 1.7 smaller than in case of using a \cite{Salpeter55} IMF.

$L_{IR}$ can be calculated by fitting the spectral energy distributions (SEDs) to appropriate templates of dusty star forming galaxies (e.g., \citealt{CE01,DH02,SK07}). In our case, W3 and W4 can be used for this purpose. We are also interested in the properties of the stellar populations in the W4BCGs,  such as stellar mass and age, which can be derived by fitting the SEDs at the bluer wavelengths to stellar population synthesis models. Therefore, we utilized LePhare (\citealt{Arn1999,Ilbert2006}), which is capable of fitting both the stellar populations and the heated dust components at the same time.

Our input SEDs were constructed using the SDSS DR7 photometry in conjunction with the unWISE photometry (as described in \S 2.2) to cover the optical to mid-IR regime. The heated dust emission was fit to the template of Sibenmorgen \& Kr\"{u}gel (2007; SK07), and the fit was confined to the W3 and W4 bands only. For the stellar component, the fit included the five SDSS bands as well as the W1 and W2 bands. LePhare treats the transition of stellar emission and heated dust emission in a consistent manner; i.e., the contribution of the dust emission template to the bands bluer than W3 (or that of the stellar emission template to the bands redder than W2), albeit small, is still considered during the simultaneous fit of the two components. The stellar component was fit to the stellar population synthesis models  of \citeauthor{BC03} (2003;~hereafter BC03). We adopted the BC03 models of solar metallicity and the Chabrier IMF, and used a series of exponentially declining  star  formation  histories
with $\tau$ ranging from 1 Myr to 20 Gyr. We note that we chose solar metallicity because we do not have any constraint on the metallicities of these objects, and the solar metallicity is the most widely adopted value throughout the literature in such case. The  models  were allowed to  be reddened by  dust  following the  Calzetti's  law (\citealt{Calz2001}), with the reddening color excess value, E(B-V), allowed to vary over three ranges: from 0 up to 0.5 mag, from 0 up to 0.3 mag, and fix to zero (i.e., no reddening). While the W4BCGs have very dusty star-forming regions, their exposed stellar populations as seen in optical-to-near-IR are not necessarily dusty. As the reddening parameter and the age of the stellar population are degenerated, we tested these three different choices to investigate the impact of different reddening values to the derived ages. Redshifts were fixed to those provided by the GMBCG Catalog.

\subsubsection{Far-IR Constraint from \textit{Herschel}}

As W3 and W4 only sample a small mid-IR window of the entire rest-frame 8-1000~$\mu$m range, it could be a concern whether they can accurately ``anchor" the fitting templates to derive $L_{IR}$. A  large number of practices in the literature have shown that one or two mid-IR bands indeed can derive $L_{IR}$ reasonably well (see e.g. \citealt{CE01,Magnelli09,Elbaz10,DH14}),  except that in the very high luminosity range such results tend to overestimate the true $L_{IR}$ (see e.g., \citealt{Elbaz11} and the references therein). In order to check how well our derivation of $L_{IR}$ above can be, we tested a few objects that also have FIR SPIRE data as described in \S 2.4, which samples the peak of the dust emissions and thus offers the most reliable derivation of $L_{IR}$ to date.
\begin{figure*}[t]
	\noindent\hspace{1.5in}{\Large \textbf{Rich}}
	\noindent\hspace{3.0in}{\Large \textbf{Poor}}
	\vspace{1pt}
	\raggedright
	
	\includegraphics[width=250pt,trim=30pt 20pt 40pt 10pt,clip]{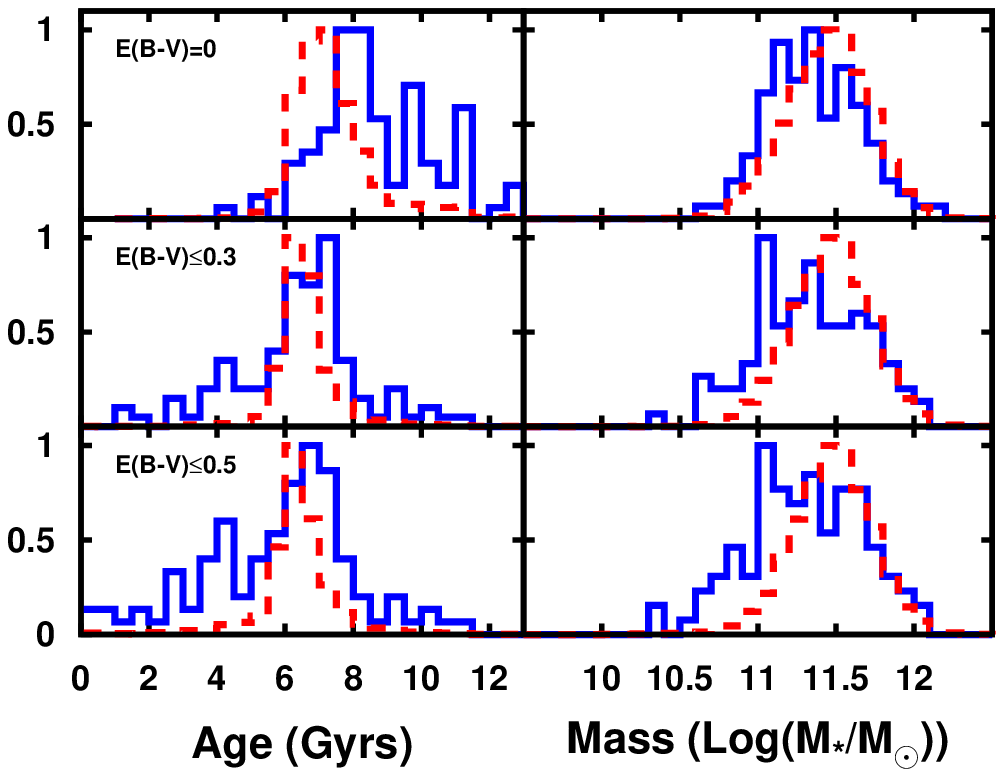}
	\includegraphics[width=250pt,trim=30pt 20pt 40pt 10pt,clip]{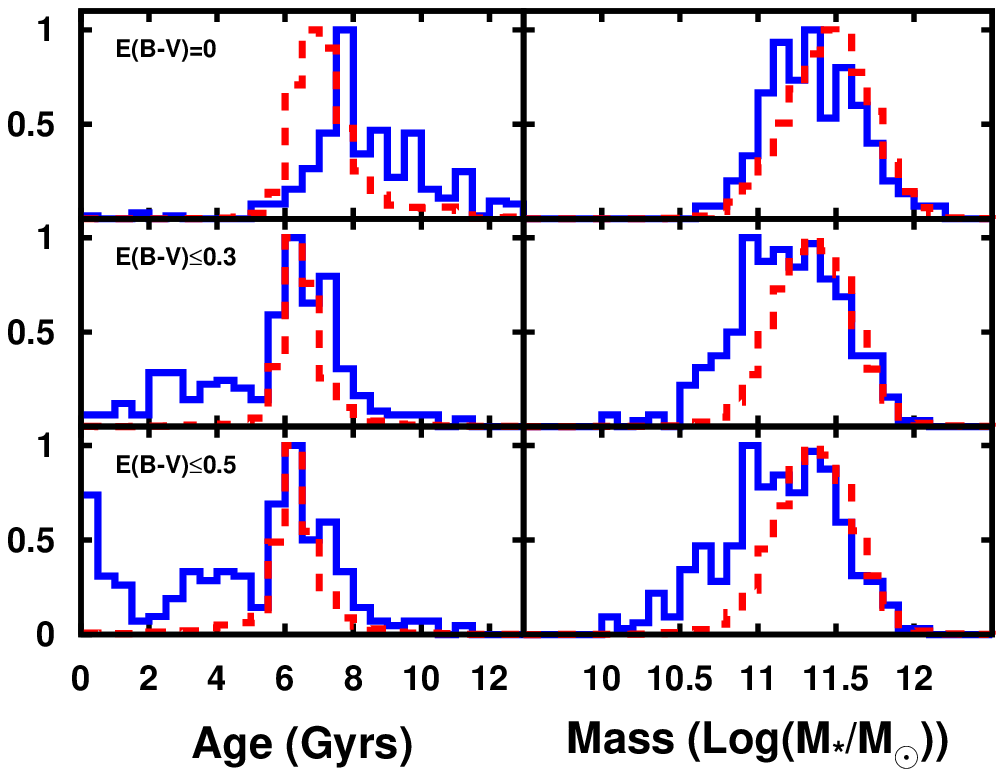}
	\caption{Left: Histogram distribution of the derived ages and masses for the ``Rich'' subset (richness $\geq$15) among W4BCGs (solid blue) and the GMBCG catalog (dashed red) for E(B-V) max values of: 0 (top), 0.3 (middle), and 0.5 (bottom). Right: Same as left except with the ``Poor'' subset (richness $<$15).}\label{fig:RP}
\end{figure*}

Following the same procedure of \cite{MY15}, we found secure SPIRE counterparts within a matching radius of 3\arcsec\, for five W4BCGs: 2 in HerS, 1 in HerMES, and 2 in H-ATLAS. A summary of the data is given in Table~\ref{tab:hersc}. We ran LePhare to fit the SEDs of these objects as before, but with the SPIRE photometry added. Fig.~\ref{fig:multi} shows their SED fitting results. For these five objects, we find that the derived median $L_{IR}$ values with and without the inclusion of the SPIRE data differ by $\sim$0.1~dex on average and 0.3~dex at most. Therefore, we believe that using W3 and W4 photometry to derive total $L_{IR}$ based on starburst templates (as in \S 4.3.1) is applicable.

\subsubsection{Results}

Applying Eq.~\ref{eq:Ken} to our sample of 389 W4BCGs results in SFRs ranging from a few to $\sim$1000 $M_\odot/yr$ (Fig.~\ref{fig:sfrhisto}). The median $L_{IR}$ is $5\times10^{11}~L_{\odot}$~(or SFR $\sim$50 $M_{\odot}/yr)$, and 27\% of the whole sample has $L_{IR}>10^{12}~L_{\odot}$ (or SFR $>$100~$M_{\odot}/yr$). The statistics largely remain the same even if we only look at W4BCG-Rs or W4BCG-Ps, or if we remove any possible AGN hosts from the sample. Obviously, the W4BCGs are not ``dead", i.e., they are not simply passively evolving like the BCG majority at low redshifts.

In addition to $L_{IR}$, SED fitting also derives stellar mass and age for each object. To check whether these W4BCGs have different stellar population properties as compared to the BCG majority, we performed SED fitting for all the non-W4BCGs from the entire GMBCG set in the same way as described in \S 4.3.1. The SEDs were based on the SDSS photometry and the unWISE photometry in W1 and W2.  Fig.~\ref{fig:RP} compares the stellar mass and the age distributions of the W4BCGs to those of the non-W4BCGs. As expected from the reddening-age degeneracy, the ages of the W4BCGs show somewhat different trends with respect to the non-W4BCGs under different choices of allowed reddening range. This is further complicated by the fact that our SED fitting templates are fixed to solar metallicity. There is also an age-metallicity degeneracy in SED fitting, in the sense that adopting a lower metallicity could result in an older age. Therefore, no definite difference in trend can be claimed regarding the age comparison. The differences in stellar mass, on the other hand, show less variation in the trends under the three choices of allowed reddening ranges. Overall speaking, we believe that the W4BCGs do not show obvious differences as compared to the non-W4BCGs in mass and age of their exposed stellar populations.

\section{Cool-Core Clusters}
As mentioned in \S  1, it has been reported in the literature that some BCGs do exhibit ongoing star formation. While their SFR triggering mechanism is unclear, they are believed to reside in ``cool-core" clusters (e.g. \citealt{MP01,ODea05,Vik07,Chen07,Santos08,Hudson10,Don15,Mac16,Mol16}). Most of these previously reported star-forming BCGs in cool-core clusters have much lower SFR as compared to the bulk of our W4BCGs, with the most notable exception of the BCG in the Phoenix cluster at $z=0.597$ (\citealt{Mac12,Tozzi15,Mitt17}), which has $SFR\sim$450 $M_\odot/yr$ (after scaling to the Chabrier IMF). In this section, we  consider whether our W4BCGs reside in cool-core clusters (hereafter ``CC clusters") as well, which can be determined by analyzing X-ray data.

\subsection{Archival Chandra Data}

We searched the {\it Chandra} archive and found that 10 W4BCGs have existing data. All observations were obtained using the ACIS instrument in either FAINT or VFAINT mode. Table~\ref{tab:Chan} summarizes these data.

\begin{table*}[t!]
	\centering
	\begin{threeparttable}
		\centering
		\caption{Summary of  Available Archival \textit{Chandra} Data}
		\begin{tabular}{l l c r r c c r r}
			\hline
			\hline
			\vspace{-0.6em}\\
			\multicolumn{1}{c}{GMBCG}& & \textit{Chandra}    &  && &Exptime& \multicolumn{1}{c}{F$_{r<40kpc}$} &\multicolumn{1}{c}{F$_{r<400kpc}$} \\
			 \multicolumn{1}{c}{Catalog Name}& ObsID &Target Name  & RA(J2000)\tnote{a} & DEC(J2000)\tnote{a}& z\tnote{b} &(ks)& \multicolumn{1}{c}{(${erg}/{s}/cm^{2}$)} & \multicolumn{1}{c}{(${erg}/{s}/cm^{2}$)}\\
			\hline
			\vspace{-0.6em}\\
			W4BCG-R &&&&&&&\\
			J027.58864-10.09181&11711&	MACS J0150.3-1005&		1:50:21.27 &-10:05:30.50&0.365	&26.8&2.27E-13 & 8.60E-13\\
			J125.25942+07.86314\tnote{c}&1647&	RXJ0821&		8:21:02.26&	7:51:47.30&$0.14^*$&	9.4&4.21E-13 & 2.12E-12\\
			--&17194&	RXJ0821.0+0752&		8:21:02.26&	7:51:47.30&$0.14^*$&	29.2&4.04E-13 &  1.79E-12\\
			--&17563&	RXJ0821.0+0752&		8:21:02.26&7:51:47.30	&$0.14^*$&	37.3&4.26E-13 & 1.78E-12\\
			J128.72875+55.57253\tnote{c}&1645&	4C55.16&		8:34:54.90&	55:34:21.10&0.241&	9.1&5.18E-13 & 1.98E-12\\
			--&4940&	4C55.16&	8:34:54.90&	55:34:21.10	&0.241&	96.0&7.41E-13& 2.14E-12\\	
			J160.18541+39.95313&1652&	ABELL 1068&		10:40:44.50	&39:57:11.30&0.138	&26.8&1.61E-12 & 5.54E-12\\
			J219.69392+06.50142&15376&	J219.69392+06.50142&		14:38:46.54&	6:30:05.10&0.403&	9.6&$<$1.81E-15 & $<$6.92E-14\\
			J355.91977+00.34170&5786&	ZwCl 2341.1+0000&		23:43:40.74&	0:20:30.10&0.261&	29.8&$<$3.72E-15 & $<$2.60E-13\\
			\\
			W4BCG-P &&&&&&&\\
			J125.63314+05.95189&12730&3C198&8:22:31.95&5:57:06.80&0.082&8.0&$<$6.38E-15&$<$3.56E-13\\
			J132.60301+37.78597&11576&6C0850+3747&8:50:24.72&37:47:09.50&$0.33^*$&39.3&1.61E-13&2.43E-13\\
			J133.71068+62.31389&16138&	RXJ085451.0+621843&		8:54:50.56&	62:18:50.00&$0.29^*$&	17.7&4.80E-13 & 5.79E-13\\
			J140.28593+45.64928&827&	3C219&	9:21:08.62&	45:38:57.40&0.174&		18.8&5.23E-13 & 7.78E-13\\
			\hline
		\end{tabular}\label{tab:Chan}
		\begin{tablenotes}
			\item[a] As quoted from the GMBCG catalog
			\item[b] Redshifts marked with an asterisk are photometric redshifts
			\item[c] J125.25942+07.86314 was observed three times, and J128.72875+55.57253 was observed twice. These data were treated separately in our follow-up analysis.
		\end{tablenotes}
	\end{threeparttable}
\end{table*}\label{tab:Chandra}

\subsection{X-ray $c_{SB}$ Parameter}

Typically, a central X-ray surface brightness excess is a good indicator of a cool core (\citealt{Fabian77}). Following this idea, \cite{Santos08} investigate the surface brightness concentration of galaxy clusters in the central region, and propose a parameter, $c_{SB}$, to distinguish between CC and non-CC clusters.

This parameter is defined as the ratio of the soft X-ray flux within 40 kpc and within 400 kpc (\citealt{Santos08}):

\begin{equation}
c_{SB} = \frac{F_{r<40kpc}}{F_{r<400kpc}}
\end{equation}

These radii are chosen because they result in the largest difference between CC and non-CC clusters. The value for $c_{SB}$ can be divided into three different regimes (\citealt{Vik07,Santos08}): non-CC ($c_{SB}<0.075$), moderate CC ($0.075<c_{SB}<0.155$), and strong CC ($c_{SB}>0.155$). Following \citet{Santos10}, we adopted the 0.5-2.0~keV band for the soft X-ray flux measurement.

\begin{figure}[t]
 	\centerline{\includegraphics[width=270pt,height=230pt,trim= 2pt 0 0 0,clip]{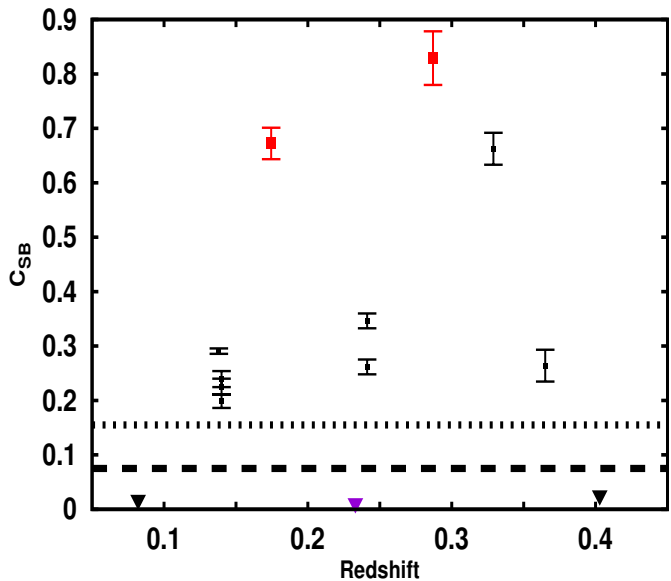}}
 	\caption{A plot of $c_{SB}$ versus redshift of the 10 W4BCGs that have archival Chandra X-ray data. The dashed line at 0.075 separates non-cool-cores from moderate cool cores. The dot-dashed line at 0.155 separates moderate cool-cores from strong cool-cores. Possible AGN hosts are color coded according to identification method:
 		BPT in blue, WISE color in red, and both in purple. The three data points at z=0.14 and the two at z=0.24 are for $c_{SB}$ derived using different observations of the same object. The triangle shows the upper limit of the object that has no X-ray detection. \label{fig:csb}}
 \end{figure}

Fig.~\ref{fig:csb} shows the values of $c_{SB}$ for all the sources with \textit{Chandra} observations. As is quite apparent, the value for $c_{SB}$ puts seven out of ten of these objects in the strong CC region, although two might have contamination due to AGN activity. However, there are three clusters not detected in X-ray at all, which we will discuss later in \S 5.4.

\subsection{X-ray Spectral Fitting}
For these seven CC clusters we further investigate their properties by carrying out X-ray spectral fitting. The spectra were fit using \texttt{XSPEC} 12.9.0 (\citealt{Arn96}) and a cooling flow model \texttt{mkcflow} (\citealt{MS88}) coupled with a single-temperature \texttt{mekal} model (\citealt{Mewe1985,Mewe1986,Kaastra92,Liedahl95}). We follow the same procedure outlined in \S 3.2 of Mol16 for the first method to find the spectral mass deposition rate. The minimum temperature for the \texttt{mkcflow} model was frozen to 0.15 keV, while the maximum temperature was frozen to 3.0 keV. The minimum temperature for the \texttt{mekal} model was set to 4.0 keV. The Galactic absorption was frozen to the value based on the radio map of \cite{Kalberla2005} at the position of the BCG. The \texttt{mkcflow} fitting outputs the mass deposition rate ($\dot{M}_{dep}$) in $M_\odot/yr$.

The fitting results are given in Table~\ref{tab:spect}. Values with no error given represent fits that did not converge (i.e., error was larger than calculated value), and we quote the upper limit based upon the 95\% confidence level. Five out of these seven have mass deposition rates falling short of the SFR estimate from $L_{IR}$. Even for the ones with $\dot{M}_{dep} > SFR$, it would require a very high efficiency ($\eta>37$--$86\%$) to convert mass into stars so that the observed SFR can be sustained by the cooling flow. Thus, it is unclear whether a possible cooling flow in these CC clusters can be responsible for the observed W4BCG star formation.

  This result is in agreement with the recent observations of some BCGs in Mol16, where the mass deposition rate was found to be an order of magnitude lower than the estimated star formation rate. Mol16 provides some possible explanations for this phenomenon, which include an origin of the gas other than the ICM, a delay between cooling and star formation, and, most likely, gas cooling out of the X-ray phase in regions much larger than those measured. However, further investigation is beyond the scope of this paper.
\subsection{Lack of Cool-Core?}
\begin{table*}[t]
	\centering
	\begin{threeparttable}
		\centering
		\caption{X-Ray Spectral Fitting Results}
		\begin{tabular}{l c c c  c c}
			\hline
			\hline
			\vspace{-0.6em}\\
			&&&Mass Deposition&Mass Deposition Rate&\\
			GMBCG Catalog Name  & \textit{Chandra} ObsID & Reduced $\chi^2$ &  Rate ($M_\odot$/yr)& Excluding 2kpc ($M_\odot$/yr)& $SFR_{L_{IR}}$ ($M_\odot$/yr)  \\
			\vspace{-0.9em}\\
			\hline
			\vspace{-0.6em}\\
			W4BCG-R &&&&\\
			J027.58864-10.09181 & 11711  &  1.6&  105.1\tnote{$\dagger$} &126.9\tnote{$\dagger$}& $215.9^{+178}_{-125}$\\
			J125.25942+07.86314 & 1647   &  1.8&  23.5$\pm$4.7 &22.8$\pm$4.7&$102.8^{+24}_{-76}$ \\
			--& 17194   &  1.9 & 23.0$\pm$3.8&20.4$\pm$3.7& $102.8^{+24}_{-76}$\\
			--& 17563 &  2.5& 30.0$\pm$3.4 &27.9$\pm$3.4& $102.8^{+24}_{-76}$\\
			J128.72875+55.57253 & 1645 &  1.7 &  20.0\tnote{$\dagger$}  &19.1\tnote{$\dagger$}& $43.3^{+18.9}_{-4}$\\
			--& 4940 &  4.0 &  17.8$\pm$8.3  &24.0$\pm$6.8& $43.3^{+18.9}_{-4}$\\
			J160.18541+39.95313 & 1652   & 3.6 &  36.7$\pm$5.3&37.4$\pm$5.2&$118.7^{+137}_{-38}$\\
			\\
			W4BCG-P &&&&\\
			J132.60301+37.78597&11576&1.7& 106.0$\pm$23.5&53.4$\pm$23.9&  $39.7^{+66}_{-14}$\\
			J133.71068+62.31389 & 16138  &   1.1& 183.1$\pm$56.3&149.2$\pm$56.9&  $157.4^{+27}_{-70}$\\
			J140.28593+45.64928 & 827  &  1.8&   --\tnote{$\ddagger$}&2.2\tnote{$\dagger$}& $135.9^{+20}_{-60}$\\
			\hline
		\end{tabular}\label{tab:spect}
		\begin{tablenotes}
			\item[\textdagger] These fitting results represent maximum values based upon 95\% confidence.
			\item[$\ddagger$] Fitting approached zero or null value.
		\end{tablenotes}
	\end{threeparttable}
\end{table*}\label{tab:spectra}
As shown in the previous section, seven of these W4BCGs are consistent with current theory by residing in cool-core clusters. However, three W4BCG do not have X-ray detection and thus show no sign of being in a cool-core, which could contradict the currently accepted picture. For these objects, the upper limit of the soft X-ray flux within a 40 kpc aperture is no larger than $2\times10^{-15}$ ${erg}\ {s}^{-1}cm^{-2}$~(for comparison, the detected sources have fluxes on the order of $10^{-13}\textendash10^{-12}$~${erg}\ {s}^{-1}cm^{-2}$). Here we discuss them briefly:

\section*{\textbf{J219.69392+06.50142}}
This W4BCG is at $z_{spec}$=$0.4029$, which is the most distant one in the X-ray sample. However, its exposure time of only 9.6 ks puts it at the shallow end of observations. Deeper X-ray observations are needed before any conclusions can be reached regarding this particular object.
 
 \section*{\textbf{J355.91977+00.34170}}
 The lack of X-ray detection at this position may be attributed to the peculiar environment that the cluster resides.~This cluster is at $z_{spec}$=$0.261$ (GMBCG gives $z_{ph}$=$0.23$), and it is merging with a nearby cluster at $z_{spec}$=$0.267$ ($z_{ph}$=$0.27$ from GMBCG) that is 5.4\arcmin\  away. The entire system is usually referred to as ZwCl 2341.1+0000, whose ICM is known to be disturbed and elongated in shape (\citealt{vW09}). While there is X-ray emission from the whole structure, the peak lies in between the two clusters and hence is offset from either BCGs. Under this circumstance, it might not be applicable to discuss the existence of a CC cluster.
 
  \section*{\textbf{J125.63314+05.95189}}
\begin{figure}[t]
	
	\plotone{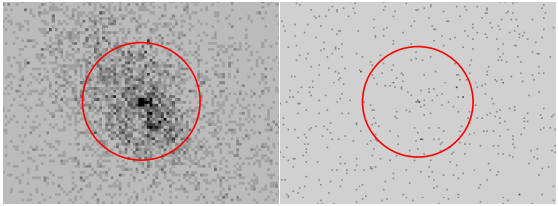}
	
	\caption{The left panel shows J128.72875+55.57253 detected in the {\it Chandra}\, data with an exposure time of 9~ks and $z_{spec}=0.24118$, while the right panel shows J125.63314+05.95189 not detected in the {\it Chandra}\, data with an exposure time of 8~ks and $z_{spec}=0.081474$. The circles are centered on the reported SDSS positions and are 40~kpc in size. \label{fig:xray}}
\end{figure}

  While the {\it Chandra}\, exposure at this position is only 8~ks, the short integration probably is not the reason for the non-detection because the object is very nearby ($z_{spec}$=$0.0815$\footnote{This BCG is at $z_{spec}$=$0.0815$ based on the SDSS DR7, however GMBCG accidentally does not use this value and keeps quoting $z_{ph}$=$0.132$ instead. Nevertheless, we verify that it indeed belongs to a cluster at $z$=$0.08$. It is surrounded by $\sim$12 red galaxies at $z_{ph}$=$0.08\pm 0.02$ that form a clear red sequence, and it is the brightest among all potential members.}). In Fig.~\ref{fig:xray}, we show that for another W4BCG with a similar exposure time yet higher redshift, there is still a clear X-ray detection. We do note that this particular W4BCG is identified as being in a low-richness or ``poor'' cluster.
  
  Due to the small sample and the aforementioned complications, we conclude that deeper X-ray observations of more W4BCGs are needed in order to put the connection of starforming BCGs and cool-core clusters on a more solid ground.

\section{Discussion}
While the common wisdom about BCGs at low redshifts is that they are quiescent galaxies, we have shown that the W4BCGs presented here are exceptions. In most cases, their W4 emissions are due to dust heated by strong star formation. Admittedly, such exceptions will not change the overall picture of BCGs because they are only a minority (W4BCGs accounting only $\sim$0.8\% among the entire GMBCG catalog). However, it is important to understand why such exceptions can happen, because this can be related to the critical question in understanding the evolution of high-mass galaxies, namely, why most high-mass galaxies have their star formation processes quenched early in time.    

We first note that the W4BCGs do not seem to have unusual environments. While it is widely believed that mergers could induce star formation, the W4BCGs are not predominantly mergers (see \S 4.2). Second, as compared to the non-W4BCGs, the W4BCGs as a whole have slightly less stellar mass and some of them can have younger ages. However, these differences are not significant and might be the results of the fitting model degeneracy rather than being real. 

Third, the cooling-flow interpretation can only explain a small fraction of W4BCGs. Among ten of them that have archival X-ray data, only seven are detected. While these seven X-ray-detected W4BCGs are consistent with being residing in cool-core clusters, five of them have their mass deposition rates (from a cooling flow model) less than their inferred SFRs.

Therefore, our investigations thus far still do not seem to be leading to a universal mechanism that can explain why W4BCGs have high SFRs. Nevertheless, there might be one clue, which is in the redshift distribution of the W4BCGs (see Fig.~\ref{fig:histos}). The high-redshift end ($z\gtrsim 0.4$) of this distribution follows that of the full GMBCG sample, which is not surprising. However, it stays relatively constant at lower redshifts, which is a feature not seen in the redshift distribution of the parent GMBCG sample. While it is still unclear how such a difference can be related to the existence of W4BCGs, it will be worth further study in the future.

\section{Conclusion}
In this paper, we present our systematic census of BCGs at low redshifts ($z<0.55$) that are still actively forming stars\footnote{During the revision of this paper, Bonavenutra et al. (2017) posted their paper on the study of star-forming BCGs, with the same main title as ours. The majority ($\sim$75\%) of the star-forming BCGs in their sample, however, are at $z>0.55$ and thus are beyond the redshift range of our W4BCGs. In this sense, the high SFRs observed in the W4BCGs are more difficult to understand because they are supposed to settle down already at such a late time of the universe.}. We use the SDSS-based GMBCG catalog, which is the largest BCG catalog to date, and identify those that have strong mid-IR emissions by their prominent detections in the W4-band (22~$\mu$m) in the WISE all-sky survey. The full catalog of these W4BCGs is presented in Table~\ref{W4table}, including their various properties as discussed in previous sections.

While some of the W4BCGs could be AGN hosts, the majority of them are not. Therefore, their strong W4 emissions should be powered by dust heating from star formation. Even for those that are possible AGN hosts, we show that their W4 emissions are still most likely due to star formation. Our W4BCGs have median SFR of $\sim$50 $M_\odot/yr$, and some have SFR as high as 500$-$1000 $M_\odot/yr$. Clearly, the W4BCGs are quite contrary to what is expected for BCGs at low redshifts, which are believed to be old, passively evolving galaxies (i.e., ``red-and-dead"). There have been a number of studies reporting some low-redshift BCGs that still have non-negligible star formation, but their SFRs are lower than what we observe among these W4BCGs and/or have smaller sample sizes. Although such actively star-forming BCGs are only a minority among all BCGs, their very existence could have important implications to the evolution of very high mass galaxies.

Our investigations so far are not able to answer why these BCGs are still actively forming stars at such a late stage. The previous studies of low-redshift BCGs that are still not completely ``dead" usually attribute the star formation triggering mechanism to the cooling flows in cool-core clusters. However, for the seven identified to be in cool-core clusters based upon X-ray data, the possible mass deposition rates due to a cooling flow fall significantly short to explain the observed SFRs, and thus the true triggering mechanism still remains a mystery. One possible clue to solve this problem could be that W4BCGs are different from the quiescent majority in their redshift distribution: their number is redshift independent as compared to the whole GMBCG sample. Further study of field galaxies will be necessary to shed new light to the understanding of this behavior.

\acknowledgments
We acknowledge the support of NASA's Astrophysics Data Analysis Program under grant number NNX15AM92G. This publication makes use of data products from the Wide-field Infrared Survey Explorer, which is a joint project of the University of California, Los Angeles, and the Jet Propulsion Laboratory/California Institute of Technology, funded by the National Aeronautics and Space Administration. We would also like to thank Zhiyuan Ma for his help with HIPE.

\bibliography{reflib}

\begin{thebibliography}{}
\expandafter\ifx\csname natexlab\endcsname\relax\def\natexlab#1{#1}\fi

\bibitem[{Arnaud(1996)}]{Arn96}
Arnaud, K.~A. 1996, in Jacoby G. H., Barnes J., eds, ASP Conf. Ser. Vol. 101,
  Astronomical Data Analysis Software and Systems V. Astron. Soc. Pac., San
  Francisco, p. 17

\bibitem[{Arnouts {et~al.}(1999)Arnouts, Cristiani, Moscardini,
  {et~al.}}]{Arn1999}
Arnouts, S., Cristiani, S., Moscardini, L., {et~al.} 1999, MNRAS, 310

\bibitem[{Assef {et~al.}(2013)Assef, Stern, Kochanek, {et~al.}}]{Assef2013}
Assef, R.~J., Stern, D., Kochanek, C.~S., {et~al.} 2013, ApJ, 772

\bibitem[{Baldwin {et~al.}(1981)Baldwin, Phillips, \& Terlevich}]{Baldwin1981}
Baldwin, J.~A., Phillips, M.~M., \& Terlevich, R. 1981, PASP, 93

\bibitem[{Brinchmann(2004)}]{Brinchmann04}
Brinchmann, J.~o. 2004, MNRAS, 351

\bibitem[{Bruzual \& Charlot(2003)}]{BC03}
Bruzual, G., \& Charlot, S. 2003, MNRAS, 344

\bibitem[{Calzetti(2001)}]{Calz2001}
Calzetti, D. 2001, PASP, 113

\bibitem[{Chabrier(2003)}]{Chab03}
Chabrier, G. 2003, PASP, 115

\bibitem[{Chary \& Elbaz(2001)}]{CE01}
Chary, R., \& Elbaz, D. 2001, ApJ, 556

\bibitem[{Chen {et~al.}(2007)}]{Chen07}
Chen, Y., {et~al.} 2007, A\&A, 466

\bibitem[{Cowie {et~al.}(1996)Cowie, Songaila, Hu, Cohen, {et~al.}}]{Cowie96}
Cowie, L.~L., Songaila, A., Hu, E.~M., Cohen, J.~G., {et~al.} 1996, AJ, 112

\bibitem[{Dale \& Helou(2002)}]{DH02}
Dale, D., \& Helou, G. 2002, ApJ, 576

\bibitem[{Dale {et~al.}(2014)Dale, Helou, {et~al.}}]{DH14}
Dale, D.~A., Helou, G., {et~al.} 2014, ApJ, 784

\bibitem[{De~Lucia {et~al.}(2006)}]{DeL06}
De~Lucia, G., {et~al.} 2006, MNRAS, 366

\bibitem[{Donahue {et~al.}(2015)}]{Don15}
Donahue, M., {et~al.} 2015, ApJ, 805

\bibitem[{Dubinski(1998)}]{Dub98}
Dubinski. 1998, ApJ, 502

\bibitem[{Eales {et~al.}(2010)}]{Eales10b}
Eales, S., {et~al.} 2010, PASP, 122

\bibitem[{Elbaz {et~al.}(2010)}]{Elbaz10}
Elbaz, D., {et~al.} 2010, A\&A, 518

\bibitem[{Elbaz {et~al.}(2011)}]{Elbaz11}
---. 2011, A\&A, 533

\bibitem[{Fabian \& Nulsen(1977)}]{Fabian77}
Fabian, A.~C., \& Nulsen, P. E.~J. 1977, MNRAS, 180

\bibitem[{Fogarty {et~al.}(2015)}]{Fog15}
Fogarty, K., {et~al.} 2015, ApJ, 813

\bibitem[{Fruscione {et~al.}(2006)}]{Fruscione06}
Fruscione, A., {et~al.} 2006, SPIE, 6270

\bibitem[{Griffin {et~al.}(2010)}]{Griffin10}
Griffin, M.~J., {et~al.} 2010, A\&A, 518

\bibitem[{Hao {et~al.}(2010)Hao, McKay, Koester, {et~al.}}]{Hao2010}
Hao, J., McKay, T.~A., Koester, B.~P., {et~al.} 2010, ApJS, 191

\bibitem[{Hudson {et~al.}(2010)Hudson, Mittal, Reiprich, {et~al.}}]{Hudson10}
Hudson, D.~S., Mittal, R., Reiprich, T.~H., {et~al.} 2010, A\&A, 513

\bibitem[{Ilbert {et~al.}(2006)Ilbert, Arnouts, McCracekn,
  {et~al.}}]{Ilbert2006}
Ilbert, O., Arnouts, S., McCracekn, H.~J., {et~al.} 2006, A\&A, 457

\bibitem[{Jarrett {et~al.}(2011)Jarrett, Cohen, {et~al.}}]{Jarrett2011}
Jarrett, T.~H., Cohen, M., {et~al.} 2011, ApJ, 735

\bibitem[{Kaastra(1992)}]{Kaastra92}
Kaastra, J. 1992, An X-Ray Spectral Code for Optically Thin Plasmas (Internal
  SRONLeiden Report, updated version 2.0)

\bibitem[{Kalberla {et~al.}(2005)Kalberla, Burton, Hartmann,
  {et~al.}}]{Kalberla2005}
Kalberla, P. M.~W., Burton, W.~B., Hartmann, D., {et~al.} 2005, A\&A, 440

\bibitem[{Kauffmann {et~al.}(2003{\natexlab{a}})Kauffmann, Heckman, Tremonti,
  Brinchmann, Charlot, White, {et~al.}}]{Kauffmann03}
Kauffmann, G., Heckman, T.~M., Tremonti, C., {et~al.} 2003{\natexlab{a}},
  MNRAS, 346

\bibitem[{Kauffmann {et~al.}(2003{\natexlab{b}})}]{Kauff03}
Kauffmann, G., {et~al.} 2003{\natexlab{b}}, MNRAS, 341

\bibitem[{Kennicutt(1998)}]{Kennicut98}
Kennicutt, R.~C. 1998, ARAA, 36

\bibitem[{Kewley {et~al.}(2001)Kewley, Dopita, Sutherland, Heisler, Trevena,
  {et~al.}}]{Kewley01}
Kewley, L.~J., Dopita, M.~A., Sutherland, R.~S., {et~al.} 2001, ApJ, 556

\bibitem[{Lang(2014)}]{Lang12014}
Lang, D. 2014, AJ, 147

\bibitem[{Lang {et~al.}(2014)Lang, Hogg, \& Schlegel}]{Lang22014}
Lang, D., Hogg, D.~W., \& Schlegel, D.~J. 2014, Arxiv e-print

\bibitem[{Liedahl {et~al.}(1995)Liedahl, Osterheld, \& Goldstein}]{Liedahl95}
Liedahl, D.~A., Osterheld, A.~L., \& Goldstein, W.~H. 1995, ApJ, 438

\bibitem[{Ma \& Yan(2015)}]{MY15}
Ma, Z., \& Yan, H. 2015, ApJ, 811

\bibitem[{Magnelii {et~al.}(2009)Magnelii, Elbaz, Chary, {et~al.}}]{Magnelli09}
Magnelii, B., Elbaz, D., Chary, R., {et~al.} 2009, A\&A, 496

\bibitem[{Mateos {et~al.}(2012)Mateos, Alonso-Herrero, Carrera,
  {et~al.}}]{Mateos2012}
Mateos, S., Alonso-Herrero, A., Carrera, F.~J., {et~al.} 2012, MNRAS, 426

\bibitem[{McDonald {et~al.}(2012)McDonald, Bayliss, Benson, {et~al.}}]{Mac12}
McDonald, M., Bayliss, M., Benson, B.~A., {et~al.} 2012, Nature, 488

\bibitem[{McDonald {et~al.}(2016)}]{Mac16}
McDonald, M., {et~al.} 2016, ApJ, 817

\bibitem[{Mewe {et~al.}(1985)Mewe, Gronenschild, \& van~den Ooord}]{Mewe1985}
Mewe, R., Gronenschild, E. H. B.~M., \& van~den Ooord, G. H.~J. 1985, A\&A, 62

\bibitem[{Mewe {et~al.}(1986)Mewe, Lemen, \& van~den Ooord}]{Mewe1986}
Mewe, R., Lemen, J.~R., \& van~den Ooord, G. H.~J. 1986, A\&A, 65

\bibitem[{Mittal {et~al.}(2017)}]{Mitt17}
Mittal, R., {et~al.} 2017, MNRAS, 465

\bibitem[{Molendi \& Pizzolato(2001)}]{MP01}
Molendi, S., \& Pizzolato, F. 2001, ApJ, 560

\bibitem[{Molendi {et~al.}(2016)Molendi, Tozzi, {et~al.}}]{Mol16}
Molendi, S., Tozzi, P., {et~al.} 2016, A\&A, 595

\bibitem[{Mushotzky \& Szymkowiak(1988)}]{MS88}
Mushotzky, R.~F., \& Szymkowiak, A. E.~a. 1988, in NATO ASIC Proc. 229: Cooling
  Flows in Clusters and Galaxies, ed. A. C. Fabian, 53-62

\bibitem[{O'Dea {et~al.}(2005)}]{ODea05}
O'Dea, C., {et~al.} 2005, ApJ, 681

\bibitem[{Oliver {et~al.}(2012)Oliver, Bock, {et~al.}}]{Oliver12}
Oliver, S., Bock, J., {et~al.} 2012, MNRAS, 424

\bibitem[{Ott(2010)}]{Ott10}
Ott, S. 2010, in ASP Conf. Ser. 434, Astronomical Data Analysis Software and
  Systems XIX, ed. Y. Mizumoto, K.-I. Morita, and M. Ohishi (San Francisco, CA:
  ASP), 139

\bibitem[{Pilbratt {et~al.}(2010)}]{Pilbratt10}
Pilbratt, G.~L., {et~al.} 2010, A\&A, 518

\bibitem[{Roseboom {et~al.}(2010)}]{Roseboom10}
Roseboom, I.~G., {et~al.} 2010, MNRAS, 409

\bibitem[{Rykoff {et~al.}(2014)}]{Rykoff14}
Rykoff, E.~S., {et~al.} 2014, ApJ, 785

\bibitem[{Salpeter(1955)}]{Salpeter55}
Salpeter, E.~E. 1955, ApJ, 121

\bibitem[{Santos {et~al.}(2008)Santos, Rosati, Tozzi, {et~al.}}]{Santos08}
Santos, J.~S., Rosati, P., Tozzi, P., {et~al.} 2008, A\&A, 483

\bibitem[{Santos {et~al.}(2010)Santos, Tozzi, Rosati, \&
  B\"{o}hringer}]{Santos10}
Santos, J.~S., Tozzi, P., Rosati, P., \& B\"{o}hringer, H. 2010, A\&A, 521

\bibitem[{Siebenmorgen \& Kr\"{u}gel(2007)}]{SK07}
Siebenmorgen, \& Kr\"{u}gel. 2007, AA, 462

\bibitem[{Smith {et~al.}(2012)}]{Smith12}
Smith, A.~J., {et~al.} 2012, MNRAS, 419

\bibitem[{Stern {et~al.}(2012)Stern, Assef, Benford, {et~al.}}]{Stern2012}
Stern, D., Assef, R.~J., Benford, D.~J., {et~al.} 2012, ApJ, 753

\bibitem[{Tozzi {et~al.}(2015)Tozzi, Gastaldello, {et~al.}}]{Tozzi15}
Tozzi, P., Gastaldello, F., {et~al.} 2015, A\&A, 580

\bibitem[{Tremonti {et~al.}(2004)}]{Tremonti04}
Tremonti, C.~A., {et~al.} 2004, ApJ, 613

\bibitem[{Valiante {et~al.}(2016)}]{Val16}
Valiante, E., {et~al.} 2016, MNRAS, 462

\bibitem[{van Weeren {et~al.}(2009)}]{vW09}
van Weeren, R.~J., {et~al.} 2009, A\&A, 506

\bibitem[{Viero {et~al.}(2013)}]{Viero13}
Viero, M.~P., {et~al.} 2013, ApJ, 772

\bibitem[{Viero {et~al.}(2014)}]{Viero14}
---. 2014, ApJS, 210

\bibitem[{Vikhlinin {et~al.}(2007)Vikhlinin, Burenin, Forman, {et~al.}}]{Vik07}
Vikhlinin, A., Burenin, R., Forman, W.~R., {et~al.} 2007, in Heating versus
  Cooling in Galaxies and Clusters of Galaxies, ed. H. B\"{o}hringer, G. W.
  Pratt, A. Finoguenov, \& P. Schuecker , 48

\bibitem[{Wang {et~al.}(2014)}]{Wang14}
Wang, L., {et~al.} 2014, MNRAS, 444

\bibitem[{Wright {et~al.}(2010)Wright, Eisenhardt, {et~al.}}]{Wright2010}
Wright, E.~L., Eisenhardt, P. R.~M., {et~al.} 2010, ApJ, 140

\bibitem[{York {et~al.}(2000)York, Adelman, {et~al.}}]{York2000}
York, D.~G., Adelman, J., {et~al.} 2000, AJ, 120

\end{thebibliography}

\clearpage
\appendix
\section{W4BCGs in redMaPPer}

While we focused on the GMBCG catalog for our search of W4BCGs, the same procedure can be applied to other cluster catalogs. One such catalog is that produced by the redMaPPer algorithm (\citealt{Rykoff14}). Like the GMBCG catalog, the redMaPPer catalog was produced by the use of a cluster finding algorithm on SDSS data. However, redMaPPer utilized SDSS DR8 photometric data and different criteria for cluster identification. The redMaPPer catalog consists of 26,111 cluster candidates covering a redshift range of $0.08<z<0.55$, roughly half the size of the GMBCG catalog despite covering a larger area and similar redshift range. To check for possible overlap between the two, we cross-matched the catalogs using a radius of \SI{400}{\arcsecond}, which corresponds to $\sim$2 Mpc at the median redshift $z=0.35$. There is a possible overlap of 14,386 cluster candidates between the redMaPPer and GMBCG catalog.

\begin{figure*}[h]
	\center
	\includegraphics[width=0.3\textwidth]{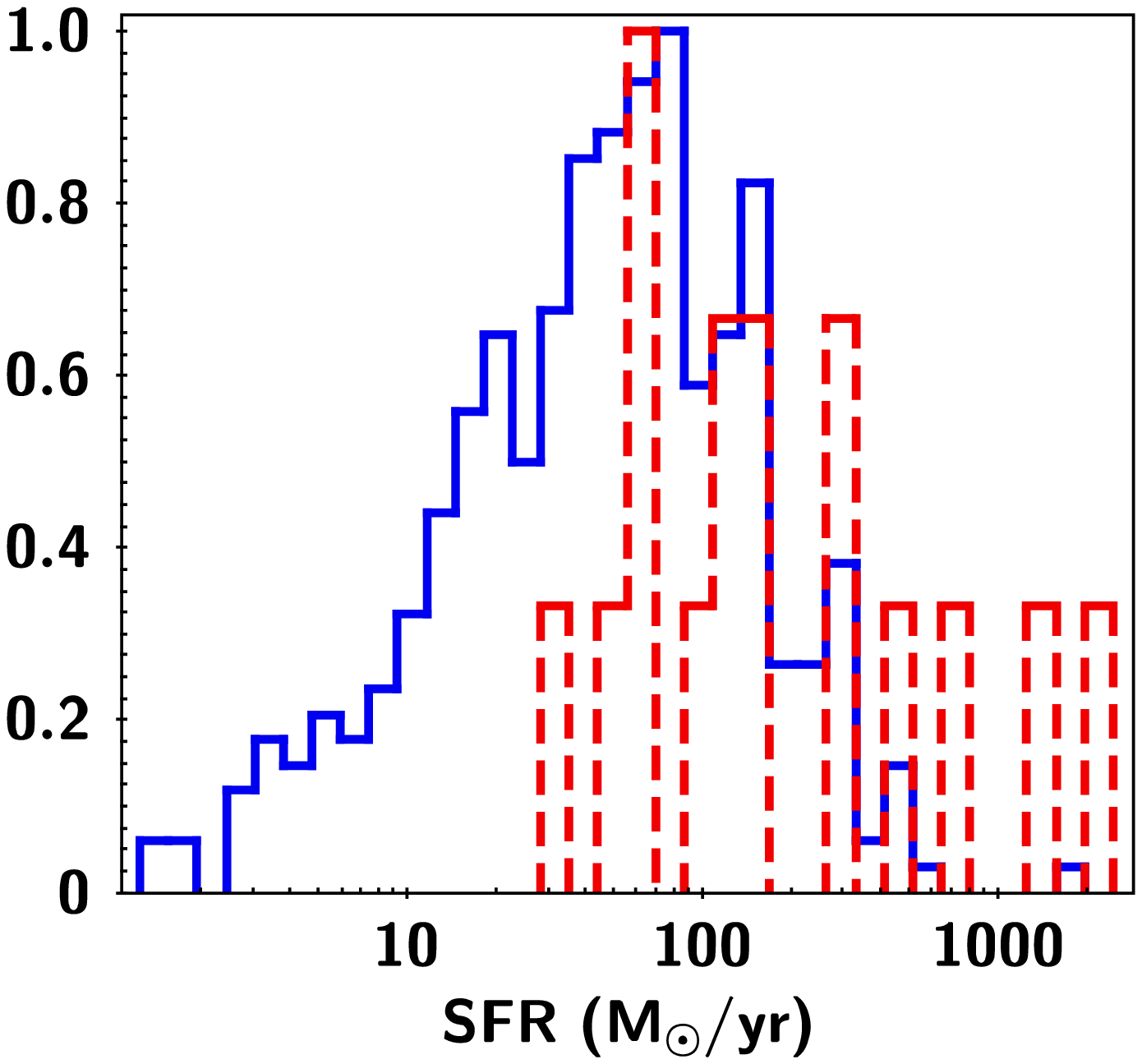}
	\includegraphics[width=0.3\textwidth]{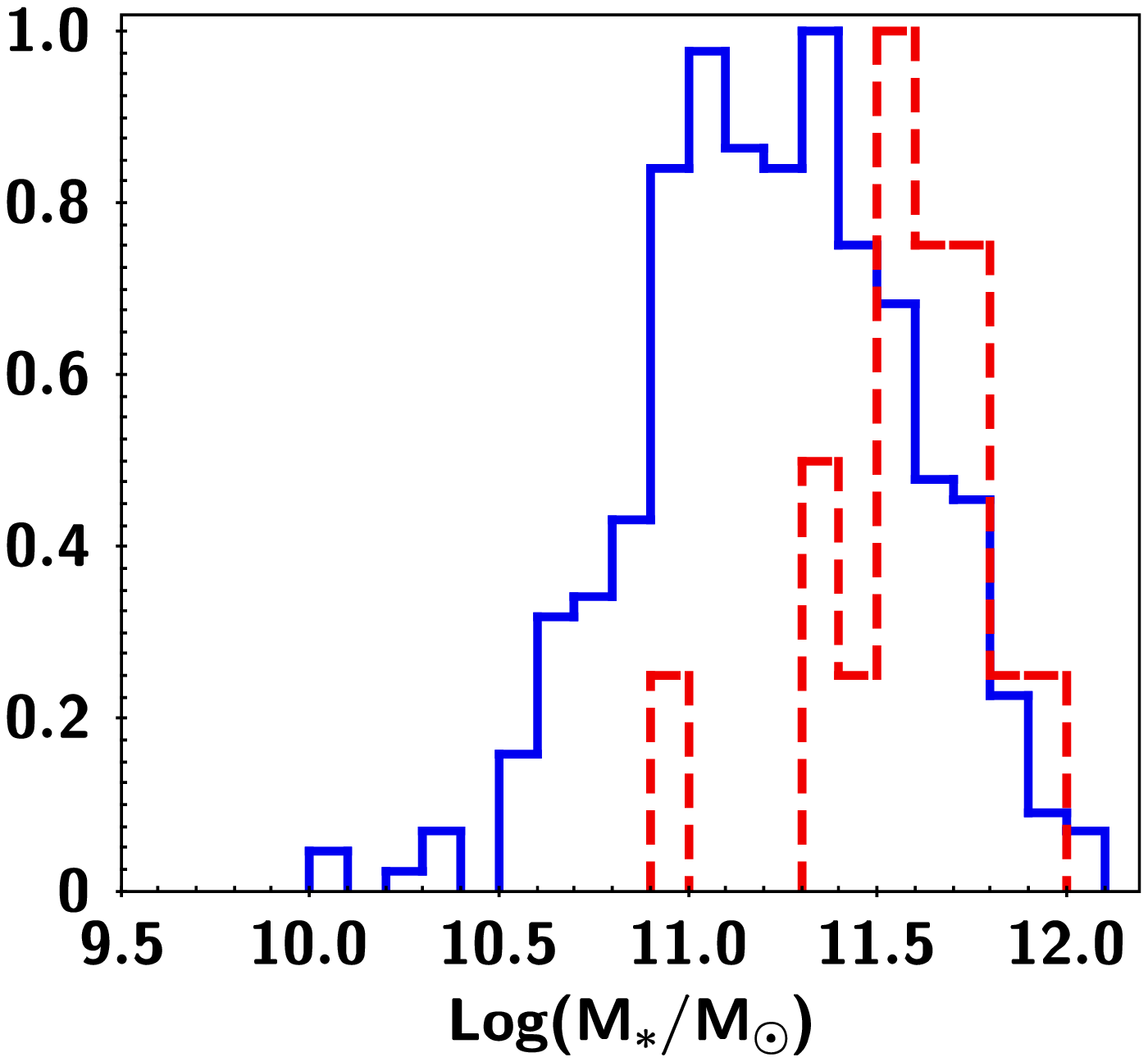}
	\includegraphics[width=0.3\textwidth]{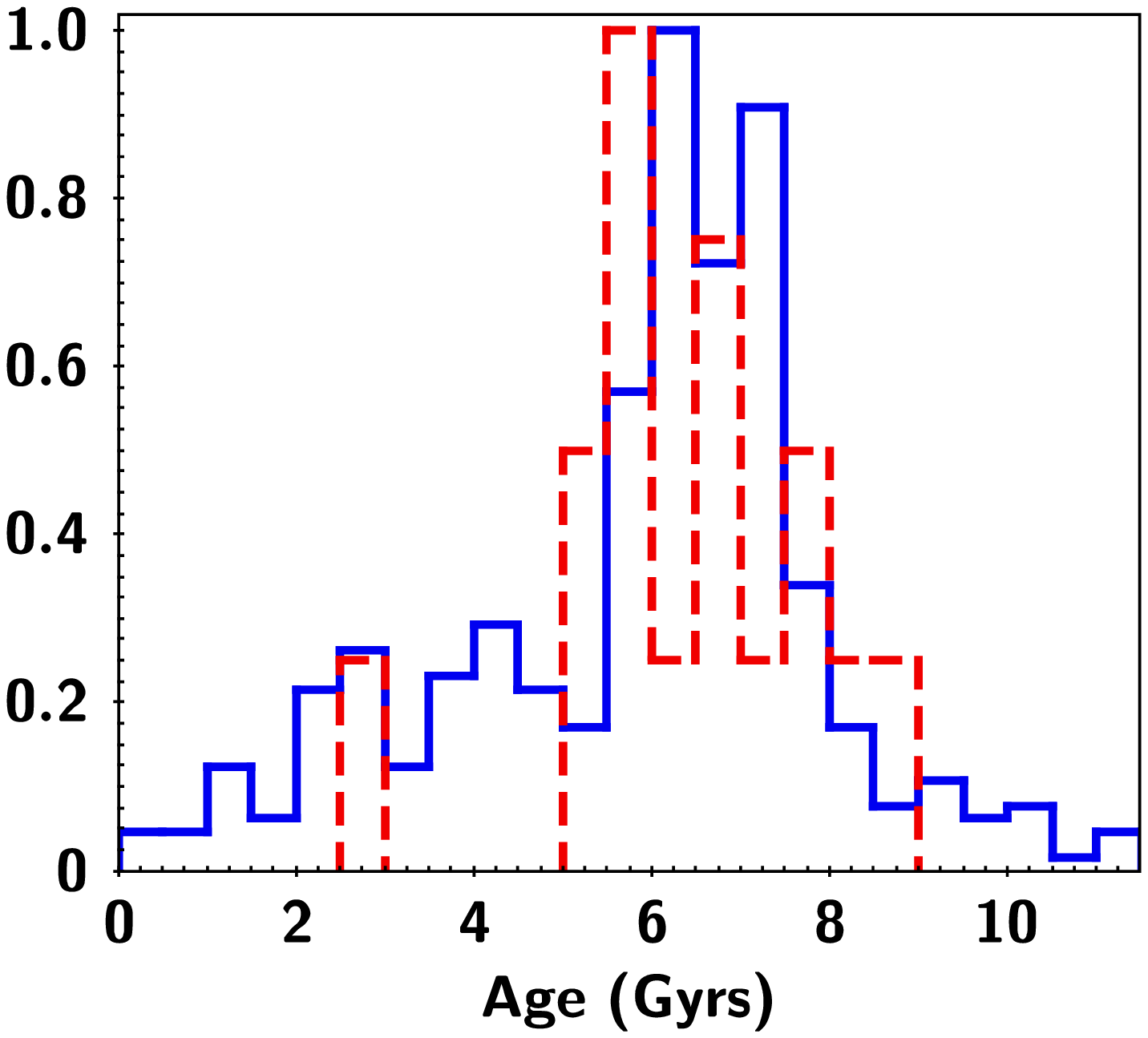}
	
	\caption{Comparison of the derived SFR  (left), stellar mass (middle), and age (right) for the W4BCGs from the GMBCG catalog (solid blue) and the redMaPPer catalog (red dashed), following the same SED fitting procedure as in \S 4.3. The fit to the stellar population is based on E(B-V) up to 0.3~mag. Each graph is normalized for easy comparison. \label{fig:red}}
\end{figure*}

Following the same procedure outlined in \S 3, we searched for BCGs in the redMaPPer catalog that had secure W4 detections. The final sample of W4BCGs in the redMaPPer consists of 16 candidates ($\sim$0.07\% of the total catalog). We preformed the same SED analysis for these objects following the procedures in \S 4.3. Their $L_{IR}$-based SFR, as well as the stellar mass and the age of their stellar populations, are shown in the histograms in Fig.~\ref{fig:red}. For comparison, we also plot the distributions of the W4BCGs from the GMBCG catalog. Despite the small number of sources, it seems that redMaPPer W4BCGs have slightly higher masses and SFRs as compared to the GMBCG sample. However, the fact remains that these W4BCGs exhibit a high amount of star formation based upon their $L_{IR}$.

	\clearpage
	
	\begin{sidewaystable*}
		
		\caption{Photometry and Derived Properties for W4BCGs}
		\label{W4table}
		
		\begin{threeparttable}
			\scalebox{0.6}{
				\begin{tabular}{llllllllllllllllllll}
					GMBCG Catalog Name  & RA(J2000)\tnote{\textdagger} & DEC(J2000)\tnote{\textdagger} & $z_{phot}$\tnote{\textdagger} & $z_{spec}$\tnote{\textdagger}&Richness\tnote{\textdagger}&u&g&r&i&z&W1&W2&W3&W4& $Log(L_{IR}/L_\odot)$& SFR($M_\odot$/yr)&Age(Gyr)&Log($M^*$/$M_\odot$)&AGN  \\
					\hline
					GMBCG J000.12121+15.71478 & 0.12121387       & 15.71477557        & 0.11$\pm$0.03 & 0.115441  & 8  & 19.42$\pm$0.07 & 17.48$\pm$0.01 & 16.48$\pm$0    & 15.98$\pm$0    & 15.61$\pm$0.01 & 15.59$\pm$0    & 15.98$\pm$0.01 & 15.3$\pm$0.06   & 14.49$\pm$0.19 & 10.94$^{+0.29}_{-0.37}$ & 8.71$^{+8.19}_{-4.97}$        & 7.4$^{+1.1}_{-3.49}$    & 10.94$^{+0.17}_{-0.06}$ & BPT  \\
					GMBCG J001.48978+15.69867 & 1.48977949       & 15.69867485        & 0.24$\pm$0.02 & 0.218624  & 10 & 21.13$\pm$0.34 & 18.88$\pm$0.02 & 17.44$\pm$0.01 & 16.87$\pm$0.01 & 16.49$\pm$0.02 & 16.33$\pm$0.01 & 16.48$\pm$0.01 & 15.57$\pm$0.08  & 14.63$\pm$0.2  & 11.56$^{+0.27}_{-0.39}$ & 36.4$^{+31.58}_{-21.49}$      & 7.5$^{+2.12}_{-1.42}$   & 11.5$^{+0.09}_{-0.06}$  & BPT  \\
					GMBCG J002.18580+00.08110 & 2.1857991        & 0.08110011         & 0.38$\pm$0.04 & 0.0       & 10 & 21.5$\pm$0.55  & 20.54$\pm$0.07 & 18.87$\pm$0.02 & 18.31$\pm$0.02 & 17.73$\pm$0.05 & 16.59$\pm$0.01 & 16.34$\pm$0.01 & 14.97$\pm$0.06  & 13.82$\pm$0.12 & 12.43$^{+0.25}_{-0.36}$ & 266.44$^{+211.86}_{-150.8}$   & 6.57$^{+1.74}_{-2.73}$  & 11.67$^{+0.1}_{-0.1}$   & WISE \\
					GMBCG J002.89046+15.21398 & 2.89045783       & 15.21398079        & 0.3$\pm$0.03  & 0.0       & 9  & 19.98$\pm$0.1  & 19.23$\pm$0.02 & 18.39$\pm$0.01 & 18.09$\pm$0.01 & 17.65$\pm$0.04 & 16.29$\pm$0    & 15.85$\pm$0.01 & 14.96$\pm$0.04  & 14.74$\pm$0.21 & 12.39$^{+0.07}_{-0.23}$ & 247.17$^{+45.65}_{-101.79}$   & 1.67$^{+0.44}_{-0.37}$  & 10.74$^{+0.06}_{-0.08}$ & WISE \\
					GMBCG J007.75699-09.61500 & 7.75699353       & -9.61499813        & 0.16$\pm$0.04 & 0.0       & 8  & 20.84$\pm$0.28 & 19.21$\pm$0.02 & 18.13$\pm$0.01 & 17.48$\pm$0.01 & 17.07$\pm$0.04 & 16.87$\pm$0.01 & 17$\pm$0.02    & 14.64$\pm$0.03  & 13.25$\pm$0.06 & 11.72$^{+0.27}_{-0.39}$ & 51.96$^{+44}_{-30.61}$        & 8.86$^{+2.22}_{-1.6}$   & 11.05$^{+0.08}_{-0.08}$ & NONE \\
					GMBCG J011.86695-00.60513 & 11.86695358      & -0.60513368        & 0.42$\pm$0.02 & 0.429752  & 24 & 22.35$\pm$1.09 & 20.66$\pm$0.08 & 18.84$\pm$0.02 & 18.11$\pm$0.02 & 17.7$\pm$0.05  & 17.27$\pm$0.02 & 17.36$\pm$0.04 & 17.5$\pm$1.26   & 13.92$\pm$0.15 & 12.1$^{+0.41}_{-0.15}$  & 125.63$^{+195.36}_{-37.47}$   & 6.45$^{+1.78}_{-2.34}$  & 11.64$^{+0.07}_{-0.11}$ & NONE \\
					GMBCG J015.05143+14.84481 & 15.05143029      & 14.8448104         & 0.38$\pm$0.07 & 0.0       & 8  & 21.72$\pm$0.3  & 20.81$\pm$0.05 & 19.35$\pm$0.02 & 18.72$\pm$0.02 & 18.2$\pm$0.05  & 16.98$\pm$0.01 & 16.7$\pm$0.01  & 15.87$\pm$0.08  & 14.72$\pm$0.17 & 12.06$^{+0.26}_{-0.35}$ & 115.74$^{+94.68}_{-63.92}$    & 7.12$^{+1.25}_{-1.62}$  & 11.53$^{+0.09}_{-0.09}$ & WISE \\
					GMBCG J027.06772+00.32915 & 27.06772378      & 0.32915166         & 0.15$\pm$0.02 & 0.0918369 & 11 & 18.16$\pm$0.03 & 16.72$\pm$0    & 15.86$\pm$0    & 15.43$\pm$0    & 15.17$\pm$0.01 & 15.09$\pm$0    & 15.27$\pm$0    & 14.81$\pm$0.03  & 14.04$\pm$0.09 & 10.85$^{+0.3}_{-0.35}$  & 7.06$^{+7.01}_{-3.92}$        & 6.5$^{+0.77}_{-0.83}$   & 11.22$^{+0.06}_{-0.27}$ & NONE \\
					GMBCG J027.24559-00.70620 & 27.24558949      & -0.7062            & 0.35$\pm$0.02 & 0.0       & 14 & 22.49$\pm$0.7  & 20.65$\pm$0.04 & 18.93$\pm$0.02 & 18.31$\pm$0.01 & 17.93$\pm$0.03 & 17.28$\pm$0.01 & 17.74$\pm$0.03 & 18.24$\pm$0.6   & 14.38$\pm$0.11 & 11.39$^{+0.29}_{-0.16}$ & 24.6$^{+22.82}_{-7.75}$       & 6.32$^{+1.86}_{-2.46}$  & 11.36$^{+0.09}_{-0.12}$ & NONE \\
					GMBCG J027.58864-10.09181 & 27.58863802      & -10.09180529       & 0.31$\pm$0.08 & 0.365     & 16 & 19.57$\pm$0.18 & 18.49$\pm$0.03 & 17.16$\pm$0.01 & 16.62$\pm$0.01 & 16.14$\pm$0.03 & 15.61$\pm$0    & 16.03$\pm$0.01 & 15.01$\pm$0.05  & 14$\pm$0.12    & 12.33$^{+0.26}_{-0.37}$ & 215.92$^{+177.99}_{-124.74}$  & 6.98$^{+1.39}_{-2.2}$   & 12.05$^{+0.08}_{-0.1}$  & BPT  \\
					GMBCG J027.86875+14.38572 & 27.86875171      & 14.38572344        & 0.35$\pm$0.1  & 0.0       & 9  & 22.07$\pm$0.28 & 21.15$\pm$0.06 & 19.86$\pm$0.03 & 19.23$\pm$0.03 & 18.94$\pm$0.06 & 17.9$\pm$0.02  & 17.97$\pm$0.04 & 16.24$\pm$0.09  & 15$\pm$0.19    & 11.83$^{+0.28}_{-0.34}$ & 67.33$^{+61.73}_{-36.2}$      & 6.38$^{+1.91}_{-2.09}$  & 11$^{+0.09}_{-0.1}$     & NONE \\
					GMBCG J029.17012-00.37641 & 29.17012063      & -0.37640939        & 0.36$\pm$0.09 & 0.0       & 12 & 21.34$\pm$0.29 & 20.63$\pm$0.06 & 19.48$\pm$0.04 & 18.96$\pm$0.03 & 18.51$\pm$0.07 & 17.78$\pm$0.02 & 17.49$\pm$0.03 & 15.9$\pm$0.07   & 14.65$\pm$0.15 & 12.01$^{+0.27}_{-0.35}$ & 103.04$^{+86.68}_{-56.68}$    & 6.08$^{+1.97}_{-2.31}$  & 11.08$^{+0.1}_{-0.11}$  & WISE \\
					GMBCG J029.56746+00.00350 & 29.56746057      & 0.0034958          & 0.42$\pm$0.02 & 0.0       & 20 & 23.48$\pm$1.2  & 21.99$\pm$0.12 & 20.26$\pm$0.04 & 19.54$\pm$0.03 & 19.09$\pm$0.09 & 18.35$\pm$0.03 & 18.84$\pm$0.09 & 18.51$\pm$0.71  & 15.01$\pm$0.19 & 11.4$^{+0.39}_{-0.18}$  & 25.04$^{+35.9}_{-8.49}$       & 6.17$^{+1.8}_{-2.63}$   & 11.09$^{+0.09}_{-0.14}$ & NONE \\
					GMBCG J029.97041-08.20800 & 29.97040719      & -8.20800187        & 0.34$\pm$0.03 & 0.346697  & 9  & 22.8$\pm$1.55  & 20.29$\pm$0.05 & 18.63$\pm$0.02 & 18$\pm$0.02    & 17.51$\pm$0.05 & 17$\pm$0.01    & 17.35$\pm$0.03 & -99$\pm$-99     & 14.59$\pm$0.16 & 11.9$^{+0.36}_{-0.27}$  & 79.67$^{+104.11}_{-36.88}$    & 6.59$^{+1.74}_{-2.45}$  & 11.52$^{+0.09}_{-0.13}$ & NONE \\
					GMBCG J034.05742-00.72531 & 34.05742358      & -0.72531111        & 0.31$\pm$0.07 & 0.0       & 15 & 19.91$\pm$0.14 & 19.12$\pm$0.02 & 18$\pm$0.01    & 17.55$\pm$0.01 & 17.14$\pm$0.03 & 16.85$\pm$0.01 & 16.93$\pm$0.02 & 15.97$\pm$0.07  & 15.07$\pm$0.2  & 11.79$^{+0.26}_{-0.38}$ & 61.12$^{+49.67}_{-35.6}$      & 6.44$^{+2.05}_{-2.26}$  & 11.3$^{+0.08}_{-0.09}$  & NONE \\
					GMBCG J038.44672-08.84924 & 38.44672304      & -8.84923564        & 0.28$\pm$0.02 & 0.0       & 18 & 22.23$\pm$1.77 & 19.02$\pm$0.03 & 17.5$\pm$0.01  & 16.96$\pm$0.01 & 16.48$\pm$0.03 & 16.36$\pm$0.01 & 16.77$\pm$0.02 & 18.2$\pm$0.74   & 14.83$\pm$0.2  & 11.12$^{+0.44}_{-0.18}$ & 13.33$^{+23.17}_{-4.51}$      & 7.58$^{+1.62}_{-1.83}$  & 11.75$^{+0.08}_{-0.13}$ & NONE \\
					GMBCG J039.77836-07.69936 & 39.77836437      & -7.69935863        & 0.39$\pm$0.07 & 0.0       & 9  & 25.05$\pm$2.79 & 20.98$\pm$0.09 & 19.68$\pm$0.05 & 18.9$\pm$0.04  & 18.48$\pm$0.09 & 17.76$\pm$0.02 & 18.09$\pm$0.05 & 16.36$\pm$0.09  & 15.1$\pm$0.19  & 11.88$^{+0.29}_{-0.32}$ & 75.16$^{+71.7}_{-38.9}$       & 6.49$^{+1.79}_{-2.68}$  & 11.28$^{+0.1}_{-0.14}$  & NONE \\
					GMBCG J044.83916+00.30161 & 44.83916234      & 0.30160525         & 0.14$\pm$0.02 & 0.13802   & 13 & 18.77$\pm$0.09 & 17.56$\pm$0.01 & 16.82$\pm$0.01 & 16.45$\pm$0.01 & 16.15$\pm$0.02 & 16.25$\pm$0.01 & 16.58$\pm$0.02 & 14.87$\pm$0.03  & 14.59$\pm$0.17 & 11.5$^{+0.11}_{-0.56}$  & 31.91$^{+9.04}_{-23.2}$       & 5.11$^{+3.71}_{-2.62}$  & 10.81$^{+0.16}_{-0.15}$ & NONE \\
					GMBCG J048.94591-07.99395 & 48.94590677      & -7.99395005        & 0.24$\pm$0.07 & 0.274238  & 12 & 19.47$\pm$0.1  & 18.45$\pm$0.01 & 17.54$\pm$0.01 & 17.2$\pm$0.01  & 16.84$\pm$0.03 & 16.39$\pm$0    & 16.17$\pm$0.01 & 14.64$\pm$0.02  & 14.15$\pm$0.12 & 12.43$^{+0.04}_{-0.19}$ & 271.33$^{+28.93}_{-95.98}$    & 2.35$^{+0.97}_{-0.62}$  & 11.07$^{+0.09}_{-0.08}$ & BOTH \\
					GMBCG J055.26483-05.56434 & 55.26483489      & -5.56434196        & 0.45$\pm$0.05 & 0.0       & 9  & 22.76$\pm$0.82 & 21.49$\pm$0.09 & 19.93$\pm$0.04 & 19.21$\pm$0.03 & 18.51$\pm$0.06 & 16.06$\pm$0    & 15.52$\pm$0    & 14.81$\pm$0.04  & 13.87$\pm$0.1  & 12.69$^{+0.23}_{-0.38}$ & 493.4$^{+351.68}_{-286.82}$   & 7.28$^{+0.96}_{-1.25}$  & 11.71$^{+0.09}_{-0.12}$ & WISE \\
					GMBCG J112.54887+42.00126 & 112.54886592     & 42.00126089        & 0.41$\pm$0.07 & 0.0       & 12 & 21.46$\pm$0.3  & 20.77$\pm$0.05 & 19.93$\pm$0.04 & 19.37$\pm$0.03 & 18.94$\pm$0.08 & 17.72$\pm$0.02 & 17.89$\pm$0.04 & 16.75$\pm$0.2   & 14.58$\pm$0.18 & 11.74$^{+0.41}_{-0.17}$ & 55.39$^{+87.11}_{-17.93}$     & 5.53$^{+1.67}_{-1.69}$  & 11.06$^{+0.09}_{-0.11}$ & NONE \\
					GMBCG J113.43436+38.87798 & 113.43435675     & 38.87798154        & 0.18$\pm$0.02 & 0.0       & 9  & 20.48$\pm$0.18 & 18.73$\pm$0.02 & 17.6$\pm$0.01  & 17.09$\pm$0.01 & 16.76$\pm$0.02 & 16.75$\pm$0.01 & 16.76$\pm$0.02 & 14.9$\pm$0.04   & 13.92$\pm$0.09 & 11.86$^{+0.11}_{-0.53}$ & 73$^{+20.16}_{-51.53}$        & 6.65$^{+2.82}_{-3.03}$  & 11.05$^{+0.12}_{-0.13}$ & NONE \\
					GMBCG J114.20870+39.33200 & 114.2086979      & 39.33200399        & 0.18$\pm$0.03 & 0.116268  & 11 & 19.09$\pm$0.08 & 17.95$\pm$0.01 & 17.16$\pm$0.01 & 16.73$\pm$0.01 & 16.52$\pm$0.02 & 16.75$\pm$0.01 & 16.66$\pm$0.02 & 14.53$\pm$0.03  & 13.14$\pm$0.05 & 11.44$^{+0.26}_{-0.39}$ & 27.27$^{+22.79}_{-16.13}$     & 6.6$^{+2.3}_{-1.77}$    & 10.67$^{+0.08}_{-0.27}$ & BPT  \\
					GMBCG J115.36870+44.40880 & 115.36869909     & 44.40880028        & 0.16$\pm$0.02 & 0.132188  & 14 & 18.43$\pm$0.06 & 16.84$\pm$0.01 & 15.83$\pm$0    & 15.38$\pm$0    & 15.07$\pm$0.01 & 15.19$\pm$0    & 15.18$\pm$0    & 13.83$\pm$0.02  & 12.86$\pm$0.04 & 11.93$^{+0.09}_{-0.58}$ & 84.7$^{+20.28}_{-62.61}$      & 7.34$^{+1.58}_{-0.88}$  & 11.63$^{+0.05}_{-0.15}$ & BPT  \\
					GMBCG J118.43924+12.64781 & 118.43924374     & 12.6478053         & 0.19$\pm$0.03 & 0.196569  & 15 & 19.79$\pm$0.07 & 18.31$\pm$0.01 & 17.27$\pm$0.01 & 16.78$\pm$0.01 & 16.58$\pm$0.02 & 16.36$\pm$0.01 & 16.58$\pm$0.02 & 14.87$\pm$0.04  & 13.55$\pm$0.08 & 11.8$^{+0.28}_{-0.38}$  & 62.85$^{+55.54}_{-36.67}$     & 4.38$^{+3.92}_{-1.95}$  & 11.1$^{+0.07}_{-0.22}$  & BPT  \\
					GMBCG J122.41201+34.92700 & 122.412007888163 & 34.9270038726458   & 0.17$\pm$0.04 & 0.0825257 & 15 & 17.59$\pm$0.01 & 16.47$\pm$0    & 15.63$\pm$0.01 & 15.23$\pm$0    & 14.9$\pm$0     & 14.87$\pm$0    & 15.17$\pm$0    & 15.12$\pm$0.05  & 14.39$\pm$0.15 & 10.41$^{+0.42}_{-0.21}$ & 2.59$^{+4.26}_{-0.99}$        & 6.31$^{+0.51}_{-0.48}$  & 11.4$^{+0.03}_{-0.03}$  & BPT  \\
					GMBCG J122.51706+41.27283 & 122.517057625448 & 41.2728302106215   & 0.2$\pm$0.04  & 0.133547  & 9  & 18.75$\pm$0.03 & 17.47$\pm$0    & 16.49$\pm$0.01 & 16.09$\pm$0    & 15.76$\pm$0.01 & 15.71$\pm$0    & 16.23$\pm$0.01 & 16.27$\pm$0.13  & 14.26$\pm$0.13 & 10.75$^{+0.42}_{-0.19}$ & 5.61$^{+9.07}_{-1.98}$        & 8.27$^{+0.81}_{-0.86}$  & 11.32$^{+0.05}_{-0.17}$ & NONE \\
					GMBCG J123.77273+07.09622 & 123.772726566836 & 7.09622106064503   & 0.12$\pm$0.03 & 0.0       & 10 & 18.02$\pm$0.02 & 16.97$\pm$0.01 & 16.41$\pm$0    & 16.02$\pm$0.01 & 15.85$\pm$0.01 & 15.74$\pm$0    & 15.98$\pm$0.01 & 13.71$\pm$0.01  & 13.19$\pm$0.06 & 11.79$^{+0.09}_{-0.59}$ & 61.8$^{+14.83}_{-45.91}$      & 2.28$^{+0.45}_{-0.4}$   & 10.37$^{+0.07}_{-0.05}$ & NONE \\
					GMBCG J124.12529+34.58306 & 124.125285691121 & 34.5830553582585   & 0.42$\pm$0.1  & 0.0       & 12 & 21.24$\pm$0.28 & 20.43$\pm$0.06 & 19.21$\pm$0.03 & 18.56$\pm$0.02 & 18.27$\pm$0.06 & 17.79$\pm$0.02 & 18.04$\pm$0.05 & 16.69$\pm$0.18  & 14.61$\pm$0.17 & 11.79$^{+0.42}_{-0.17}$ & 61.83$^{+99.12}_{-20.12}$     & 6.74$^{+1.47}_{-2.58}$  & 11.29$^{+0.08}_{-0.1}$  & NONE \\
					GMBCG J125.03628+56.72644 & 125.036284701265 & 56.726436006056    & 0.12$\pm$0.03 & 0.0807631 & 9  & 17.41$\pm$0.02 & 16.12$\pm$0    & 15.39$\pm$0    & 14.95$\pm$0    & 14.69$\pm$0    & 14.49$\pm$0    & 14.79$\pm$0    & 12.6$\pm$0      & 11.73$\pm$0.01 & 11.3$^{+0.03}_{-0.03}$  & 19.94$^{+1.29}_{-1.21}$       & 5.76$^{+2.18}_{-1.59}$  & 11.01$^{+0.05}_{-0.05}$ & NONE \\
					GMBCG J125.25942+07.86314 & 125.25942181613  & 7.86313797474906   & 0.14$\pm$0.02 & 0.0       & 24 & 17.98$\pm$0.03 & 16.71$\pm$0    & 15.94$\pm$0    & 15.48$\pm$0    & 15.23$\pm$0.01 & 15.05$\pm$0    & 15.44$\pm$0.01 & 13.83$\pm$0.02  & 12.8$\pm$0.04  & 12.01$^{+0.09}_{-0.58}$ & 102.85$^{+23.57}_{-75.82}$    & 6.02$^{+1.85}_{-1.7}$   & 11.24$^{+0.15}_{-0.07}$ & NONE \\
					GMBCG J125.33837+16.12444 & 125.338367115871 & 16.1244373982432   & 0.12$\pm$0.02 & 0.126844  & 8  & 19.01$\pm$0.05 & 17.79$\pm$0.01 & 17.1$\pm$0.01  & 16.7$\pm$0.01  & 16.46$\pm$0.01 & 16.47$\pm$0.01 & 16.84$\pm$0.02 & 14.88$\pm$0.04  & 14.16$\pm$0.13 & 11.34$^{+0.24}_{-0.42}$ & 22.05$^{+16.4}_{-13.66}$      & 3.87$^{+0.81}_{-1.36}$  & 10.55$^{+0.06}_{-0.05}$ & NONE \\
					GMBCG J126.54276+15.86042 & 126.542762757289 & 15.8604177369356   & 0.19$\pm$0.03 & 0.0       & 13 & 18.85$\pm$0.03 & 18.4$\pm$0.01  & 17.64$\pm$0.01 & 17.12$\pm$0.01 & 17$\pm$0.01    & 16.41$\pm$0.01 & 16.32$\pm$0.01 & 15.36$\pm$0.06  & 14.32$\pm$0.15 & 11.55$^{+0.26}_{-0.4}$  & 35.3$^{+28.39}_{-21.33}$      & 8.81$^{+1.02}_{-0.93}$  & 11.06$^{+0.04}_{-0.04}$ & NONE \\
					GMBCG J126.74500+53.21243 & 126.744995027599 & 53.2124278025339   & 0.11$\pm$0.01 & 0.117597  & 18 & 18.17$\pm$0.06 & 16.14$\pm$0    & 15.14$\pm$0.01 & 14.69$\pm$0    & 14.33$\pm$0    & 14.37$\pm$0    & 14.85$\pm$0.01 & 14.45$\pm$0.04  & 13.87$\pm$0.14 & 11.12$^{+0.31}_{-0.33}$ & 13.05$^{+13.74}_{-7.01}$      & 4.66$^{+0.61}_{-0.45}$  & 11.32$^{+0.05}_{-0.04}$ & BPT  \\
					GMBCG J127.03216+08.62914 & 127.032164094954 & 8.62914486906813   & 0.43$\pm$0.1  & 0.0       & 15 & 21.59$\pm$0.17 & 20.91$\pm$0.04 & 19.8$\pm$0.02  & 19.15$\pm$0.02 & 18.75$\pm$0.04 & 17.02$\pm$0.01 & 16.71$\pm$0.01 & 15.37$\pm$0.08  & 14.17$\pm$0.17 & 12.38$^{+0.29}_{-0.31}$ & 242.49$^{+225.78}_{-124.65}$  & 7.08$^{+1.07}_{-1.5}$   & 11.35$^{+0.06}_{-0.07}$ & WISE \\
					GMBCG J127.52713+14.76496 & 127.527132473944 & 14.7649592089859   & 0.3$\pm$0.04  & 0.0       & 9  & 19.14$\pm$0.06 & 18.69$\pm$0.02 & 18$\pm$0.01    & 17.64$\pm$0.01 & 17.28$\pm$0.03 & 16.96$\pm$0.01 & 17.09$\pm$0.03 & 15.38$\pm$0.06  & 14.71$\pm$0.2  & 12.02$^{+0.24}_{-0.41}$ & 105.49$^{+76.27}_{-64.03}$    & 0.96$^{+1.38}_{-0.16}$  & 10.74$^{+0.09}_{-0.06}$ & NONE \\
					GMBCG J128.45608+47.50471 & 128.456075979887 & 47.5047107411214   & 0.46$\pm$0.12 & 0.0       & 18 & 20.89$\pm$0.19 & 20.47$\pm$0.05 & 19.31$\pm$0.03 & 18.61$\pm$0.02 & 18.26$\pm$0.06 & 16.32$\pm$0    & 16.04$\pm$0.01 & 15.32$\pm$0.05  & 14.25$\pm$0.12 & 12.49$^{+0.26}_{-0.35}$ & 312.54$^{+255.01}_{-172.13}$  & 7.06$^{+1.12}_{-1.8}$   & 11.71$^{+0.06}_{-0.09}$ & WISE \\
					GMBCG J128.51842+13.71544 & 128.518423629677 & 13.71543671725     & 0.19$\pm$0.02 & 0.179865  & 10 & 20.2$\pm$0.15  & 18.49$\pm$0.01 & 17.32$\pm$0.01 & 16.77$\pm$0.01 & 16.44$\pm$0.01 & 16.24$\pm$0.01 & 16.26$\pm$0.01 & 15.85$\pm$0.2   & 14.48$\pm$0.21 & 11.13$^{+0.43}_{-0.21}$ & 13.59$^{+23.32}_{-5.3}$       & 7.31$^{+2.98}_{-3.43}$  & 11.31$^{+0.1}_{-0.05}$  & BPT  \\
					GMBCG J128.72875+55.57253 & 128.728748440917 & 55.5725303361728   & 0.2$\pm$0.03  & 0.241181  & 17 & 19.63$\pm$0.18 & 17.93$\pm$0.01 & 16.73$\pm$0.01 & 16.11$\pm$0.01 & 15.86$\pm$0.02 & 15.76$\pm$0    & 16.12$\pm$0.01 & 15.13$\pm$0.05  & 12.96$\pm$0.04 & 11.64$^{+0.16}_{-0.04}$ & 43.32$^{+18.89}_{-3.67}$      & 7.95$^{+0.69}_{-0.63}$  & 11.85$^{+0.04}_{-0.04}$ & BPT  \\
					GMBCG J129.17324+03.01820 & 129.173244962775 & 3.01820149333377   & 0.45$\pm$0.03 & 0.0       & 13 & 22.33$\pm$0.94 & 21.32$\pm$0.09 & 19.71$\pm$0.03 & 18.92$\pm$0.02 & 18.41$\pm$0.08 & 18.04$\pm$0.03 & 18.84$\pm$0.11 & 19.6$\pm$2.74   & 14.09$\pm$0.11 & 11.68$^{+0.37}_{-0.17}$ & 48.31$^{+64.23}_{-15.4}$      & 6$^{+1.77}_{-2.44}$     & 11.3$^{+0.07}_{-0.15}$  & NONE \\
					GMBCG J130.08931+17.24245 & 130.089305109374 & 17.242450327423    & 0.1$\pm$0.03  & 0.059579  & 8  & 17.8$\pm$0.02  & 16.61$\pm$0    & 15.98$\pm$0    & 15.59$\pm$0    & 15.38$\pm$0.01 & 15.2$\pm$0     & 15.5$\pm$0.01  & 14.01$\pm$0.03  & 14.04$\pm$0.16 & 10.4$^{+0.04}_{-0.04}$  & 2.53$^{+0.27}_{-0.2}$         & 4.73$^{+0.7}_{-0.79}$   & 10$^{+0.04}_{-0.04}$    & BPT  \\
					GMBCG J130.54583+59.92378 & 130.545833209581 & 59.9237839102679   & 0.15$\pm$0.03 & 0.12812   & 12 & 18.32$\pm$0.03 & 16.84$\pm$0    & 15.96$\pm$0    & 15.5$\pm$0     & 15.21$\pm$0.01 & 14.99$\pm$0    & 15.34$\pm$0    & 13.61$\pm$0.01  & 13.47$\pm$0.05 & 11.77$^{+0.05}_{-0.61}$ & 59.31$^{+7.82}_{-44.72}$      & 2.57$^{+0.29}_{-0.23}$  & 11.39$^{+0.04}_{-0.14}$ & NONE \\
					GMBCG J130.97171+09.84038 & 130.971712696306 & 9.84037989028271   & 0.34$\pm$0.08 & 0.0       & 9  & 20.98$\pm$0.18 & 20.14$\pm$0.03 & 18.61$\pm$0.01 & 18.05$\pm$0.01 & 17.62$\pm$0.02 & 17.28$\pm$0.01 & 17.61$\pm$0.03 & -99$\pm$-99     & 14.64$\pm$0.17 & 11.88$^{+0.35}_{-0.29}$ & 75.37$^{+92.43}_{-36.29}$     & 7.1$^{+1.64}_{-2.01}$   & 11.43$^{+0.09}_{-0.09}$ & NONE \\
					GMBCG J131.94463+23.03120 & 131.944634932858 & 23.0311980313457   & 0.35$\pm$0.11 & 0.0       & 14 & 21.85$\pm$0.45 & 20.59$\pm$0.05 & 19.39$\pm$0.03 & 18.66$\pm$0.02 & 18.2$\pm$0.06  & 16.85$\pm$0.01 & 16.96$\pm$0.02 & 15.61$\pm$0.25  & 14.14$\pm$0.15 & 11.97$^{+0.44}_{-0.18}$ & 92.47$^{+162.15}_{-31.38}$    & 7.5$^{+1.08}_{-1.49}$   & 11.42$^{+0.07}_{-0.09}$ & NONE \\
					GMBCG J132.09946+43.80267 & 132.09945628613  & 43.8026657148125   & 0.16$\pm$0.01 & 0.152366  & 11 & 19.67$\pm$0.09 & 17.52$\pm$0    & 16.39$\pm$0.01 & 15.92$\pm$0    & 15.64$\pm$0.01 & 15.65$\pm$0    & 16.12$\pm$0.01 & 17.73$\pm$0.47  & 14.41$\pm$0.16 & 10.77$^{+0.43}_{-0.19}$ & 5.82$^{+9.73}_{-2.08}$        & 5.51$^{+0.66}_{-2.04}$  & 11.41$^{+0.05}_{-0.08}$ & BPT  \\
					GMBCG J132.61454+52.27889 & 132.614538186751 & 52.2788861995261   & 0.41$\pm$0.05 & 0.0       & 11 & 22.03$\pm$0.34 & 21.22$\pm$0.06 & 19.7$\pm$0.03  & 19.04$\pm$0.02 & 18.62$\pm$0.05 & 17.3$\pm$0.01  & 16.61$\pm$0.01 & 15.27$\pm$0.04  & 14.29$\pm$0.12 & 12.4$^{+0.24}_{-0.38}$  & 252.29$^{+189.69}_{-146.61}$  & 6.51$^{+1.57}_{-1.99}$  & 11.45$^{+0.08}_{-0.12}$ & WISE \\
					GMBCG J132.96057+39.82724 & 132.960567769331 & 39.8272393226648   & 0.35$\pm$0.03 & 0.345035  & 22 & 21.58$\pm$0.53 & 19.73$\pm$0.03 & 18.06$\pm$0.02 & 17.42$\pm$0.01 & 17.07$\pm$0.02 & 16.38$\pm$0.01 & 16.93$\pm$0.03 & 16.06$\pm$0.13  & 14.61$\pm$0.21 & 11.72$^{+0.42}_{-0.19}$ & 53.04$^{+88.02}_{-18.62}$     & 6.41$^{+1.84}_{-2.44}$  & 11.7$^{+0.09}_{-0.1}$   & NONE \\
					GMBCG J133.31805+41.40923 & 133.318053383366 & 41.4092349195037   & 0.17$\pm$0.03 & 0.133226  & 10 & 19.17$\pm$0.04 & 18.02$\pm$0    & 17.22$\pm$0.01 & 16.75$\pm$0    & 16.56$\pm$0.01 & 16.52$\pm$0.01 & 16.86$\pm$0.02 & 15.58$\pm$0.06  & 14.36$\pm$0.14 & 11.11$^{+0.27}_{-0.38}$ & 12.76$^{+11.12}_{-7.42}$      & 5.75$^{+1.11}_{-2.67}$  & 10.94$^{+0.08}_{-0.15}$ & BPT  \\
					GMBCG J133.71068+62.31389 & 133.710675780926 & 62.3138892763371   & 0.29$\pm$0.05 & 0.0       & 11 & 19.09$\pm$0.04 & 18.65$\pm$0.01 & 17.99$\pm$0.01 & 17.62$\pm$0.01 & 17.31$\pm$0.01 & 16.19$\pm$0    & 15.95$\pm$0    & 15.31$\pm$0.04  & 14.64$\pm$0.13 & 12.2$^{+0.07}_{-0.26}$  & 157.36$^{+27.35}_{-70.04}$    & 0.96$^{+0.12}_{-0.1}$   & 10.74$^{+0.05}_{-0.04}$ & WISE \\
					GMBCG J134.10410+04.16905 & 134.104103659888 & 4.16905462271393   & 0.28$\pm$0.03 & 0.0       & 10 & 23.72$\pm$2.36 & 20.13$\pm$0.03 & 18.66$\pm$0.02 & 18.07$\pm$0.01 & 17.66$\pm$0.03 & 17.34$\pm$0.02 & 17.78$\pm$0.04 & 18.58$\pm$1.26  & 14.58$\pm$0.2  & 11.1$^{+0.43}_{-0.18}$  & 12.7$^{+21.39}_{-4.24}$       & 7.4$^{+1.37}_{-1.95}$   & 11.31$^{+0.08}_{-0.09}$ & NONE \\
					GMBCG J134.81077+14.43933 & 134.810766758989 & 14.4393323466222   & 0.41$\pm$0.06 & 0.0       & 10 & 22.86$\pm$0.67 & 21.75$\pm$0.09 & 20.22$\pm$0.04 & 19.48$\pm$0.03 & 19.06$\pm$0.07 & 18.03$\pm$0.02 & 18.07$\pm$0.05 & 15.87$\pm$0.08  & 14.47$\pm$0.14 & 12.01$^{+0.41}_{-0.18}$ & 103.35$^{+164.57}_{-35.13}$   & 6.77$^{+1.41}_{-2.81}$  & 11.2$^{+0.09}_{-0.1}$   & NONE \\
					GMBCG J135.66011+17.63098 & 135.660109294574 & 17.6309768188307   & 0.2$\pm$0.02  & 0.164037  & 8  & 19.3$\pm$0.06  & 17.9$\pm$0.01  & 16.88$\pm$0.01 & 16.37$\pm$0.01 & 16.09$\pm$0.01 & 15.6$\pm$0     & 15.63$\pm$0.01 & 15.05$\pm$0.06  & 14.03$\pm$0.13 & 11.48$^{+0.27}_{-0.39}$ & 30.28$^{+26.29}_{-18.08}$     & 4.02$^{+2.45}_{-1.54}$  & 11.33$^{+0.1}_{-0.07}$  & BPT  \\
					GMBCG J135.73803+27.02423 & 135.738026696358 & 27.0242347317556   & 0.42$\pm$0.03 & 0.0       & 30 & 21.38$\pm$0.45 & 20.7$\pm$0.07  & 19.07$\pm$0.03 & 18.34$\pm$0.02 & 17.93$\pm$0.05 & 16.59$\pm$0.01 & 16.54$\pm$0.02 & 15.52$\pm$0.07  & 14.66$\pm$0.19 & 12.29$^{+0.27}_{-0.35}$ & 192.97$^{+163.23}_{-107.31}$  & 6.66$^{+1.53}_{-2.81}$  & 11.76$^{+0.09}_{-0.1}$  & NONE \\
					GMBCG J136.04140+34.22219 & 136.041396626264 & 34.2221866685943   & 0.48$\pm$0.09 & 0.0       & 15 & 21.73$\pm$0.39 & 21.11$\pm$0.09 & 19.93$\pm$0.05 & 19.19$\pm$0.04 & 18.81$\pm$0.09 & 17.73$\pm$0.02 & 17.87$\pm$0.04 & 15.74$\pm$0.08  & 14.75$\pm$0.22 & 12.37$^{+0.27}_{-0.33}$ & 232.92$^{+195.93}_{-123.7}$   & 5.86$^{+1.78}_{-2.49}$  & 11.3$^{+0.1}_{-0.12}$   & NONE \\
					GMBCG J136.13329+57.30211 & 136.133290986086 & 57.3021123351345   & 0.28$\pm$0.03 & 0.0       & 8  & 19.49$\pm$0.05 & 18.8$\pm$0.01  & 17.97$\pm$0.01 & 17.57$\pm$0.01 & 17.2$\pm$0.02  & 16.77$\pm$0.01 & 16.82$\pm$0.01 & 15.27$\pm$0.04  & 14.94$\pm$0.21 & 12.12$^{+0.08}_{-0.44}$ & 133.08$^{+26.81}_{-84.79}$    & 2.85$^{+1.14}_{-0.54}$  & 10.9$^{+0.09}_{-0.05}$  & NONE \\
					GMBCG J137.17288+61.00169 & 137.172881824142 & 61.0016911370256   & 0.35$\pm$0.05 & 0.0       & 12 & 20.46$\pm$0.11 & 19.89$\pm$0.02 & 18.87$\pm$0.02 & 18.39$\pm$0.01 & 17.91$\pm$0.04 & 16.42$\pm$0    & 16.12$\pm$0.01 & 15.62$\pm$0.05  & 14.39$\pm$0.11 & 12.07$^{+0.28}_{-0.35}$ & 117.95$^{+106.28}_{-64.74}$   & 6.39$^{+1.46}_{-1.35}$  & 11.21$^{+0.08}_{-0.08}$ & WISE \\
					GMBCG J137.23314+31.16090 & 137.233141058342 & 31.1608957019611   & 0.31$\pm$0.02 & 0.313533  & 16 & 22.6$\pm$0.78  & 20.03$\pm$0.04 & 18.47$\pm$0.01 & 17.84$\pm$0.01 & 17.54$\pm$0.03 & 17.03$\pm$0.01 & 17.48$\pm$0.04 & 19.63$\pm$3.29  & 14.41$\pm$0.17 & 11.22$^{+0.42}_{-0.17}$ & 16.63$^{+27.15}_{-5.51}$      & 6.86$^{+1.83}_{-2.3}$   & 11.39$^{+0.07}_{-0.09}$ & NONE \\
					GMBCG J137.45253+10.94288 & 137.452534451653 & 10.9428802941904   & 0.16$\pm$0.03 & 0.164306  & 19 & 18.73$\pm$0.09 & 17.32$\pm$0.01 & 16.41$\pm$0.01 & 15.98$\pm$0.01 & 15.81$\pm$0.02 & 15.72$\pm$0.01 & 15.85$\pm$0.01 & 14.79$\pm$0.06  & 13.96$\pm$0.14 & 11.6$^{+0.25}_{-0.4}$   & 39.98$^{+31.57}_{-24.18}$     & 3.77$^{+3.14}_{-2.53}$  & 11.22$^{+0.13}_{-0.12}$ & BPT  \\
					GMBCG J137.63596+41.04753 & 137.635964116445 & 41.0475327184696   & 0.48$\pm$0.08 & 0.0       & 20 & 21.35$\pm$0.22 & 20.48$\pm$0.03 & 19.22$\pm$0.02 & 18.48$\pm$0.01 & 18.06$\pm$0.03 & 16.83$\pm$0.01 & 16.52$\pm$0.01 & 15.65$\pm$0.07  & 14.76$\pm$0.18 & 12.36$^{+0.25}_{-0.35}$ & 229.77$^{+179.2}_{-125.97}$   & 6.04$^{+1.61}_{-1.92}$  & 11.7$^{+0.06}_{-0.1}$   & WISE \\
					GMBCG J137.99779+59.50252 & 137.997790001713 & 59.5025206498211   & 0.23$\pm$0.04 & 0.0       & 9  & 17.69$\pm$0.01 & 17.43$\pm$0.01 & 16.68$\pm$0.01 & 16.15$\pm$0    & 15.93$\pm$0.01 & 15.45$\pm$0    & 15.53$\pm$0.01 & 14.6$\pm$0.03   & 13.9$\pm$0.1   & 12.21$^{+0.07}_{-0.27}$ & 162.07$^{+29.05}_{-74.89}$    & 10                      & 11.58                   & NONE \\
					GMBCG J138.72789+05.55243 & 138.727889276118 & 5.55242909843523   & 0.18$\pm$0.03 & 0.195994  & 17 & 18.63$\pm$0.05 & 17.59$\pm$0.01 & 16.89$\pm$0.01 & 16.5$\pm$0     & 16.31$\pm$0.02 & 16.27$\pm$0.01 & 16.56$\pm$0.02 & 14.61$\pm$0.03  & 14.38$\pm$0.14 & 12.05$^{+0.06}_{-0.21}$ & 112.43$^{+17.34}_{-42.51}$    & 3.82$^{+4.22}_{-1.53}$  & 10.88$^{+0.11}_{-0.1}$  & NONE \\
					GMBCG J139.21642+52.64121 & 139.21641628926  & 52.6412142213301   & 0.27$\pm$0.06 & 0.190389  & 16 & 18.39$\pm$0.02 & 17.57$\pm$0    & 16.68$\pm$0.01 & 16.32$\pm$0    & 15.98$\pm$0.01 & 15.84$\pm$0    & 15.92$\pm$0.01 & 16.17$\pm$0.1   & 14.53$\pm$0.15 & 11.06$^{+0.45}_{-0.18}$ & 11.57$^{+21.27}_{-3.94}$      & 6.59$^{+2.2}_{-1.88}$   & 11.41$^{+0.06}_{-0.06}$ & BPT  \\
					GMBCG J140.28593+45.64928 & 140.285925333149 & 45.649278453841    & 0.21$\pm$0.03 & 0.174485  & 8  & 19.11$\pm$0.05 & 17.82$\pm$0    & 16.69$\pm$0.01 & 16.22$\pm$0    & 15.95$\pm$0.01 & 15.15$\pm$0    & 14.94$\pm$0    & 14.19$\pm$0.02  & 13.27$\pm$0.06 & 12.13$^{+0.06}_{-0.25}$ & 135.86$^{+20.02}_{-60.16}$    & 7.88$^{+0.73}_{-0.67}$  & 11.6$^{+0.04}_{-0.04}$  & WISE \\
					GMBCG J140.82888+29.76941 & 140.828877469209 & 29.7694076793216   & 0.39$\pm$0.07 & 0.0       & 11 & 20.69$\pm$0.17 & 20.12$\pm$0.04 & 19.48$\pm$0.03 & 18.99$\pm$0.03 & 18.25$\pm$0.04 & 17.93$\pm$0.02 & 18.14$\pm$0.06 & 16$\pm$0.1      & 14.52$\pm$0.2  & 11.92$^{+0.41}_{-0.19}$ & 82.28$^{+129.07}_{-29.02}$    & 5.64$^{+1.93}_{-3.29}$  & 10.88$^{+0.11}_{-0.16}$ & NONE \\
					GMBCG J141.59072+62.46469 & 141.590718896807 & 62.4646931596678   & 0.11$\pm$0.02 & 0.12634   & 9  & 18.93$\pm$0.07 & 17.32$\pm$0.01 & 16.45$\pm$0    & 16.04$\pm$0    & 15.75$\pm$0.01 & 15.78$\pm$0    & 16.23$\pm$0.01 & 15.4$\pm$0.05   & 14.95$\pm$0.21 & 10.96$^{+0.27}_{-0.4}$  & 9.19$^{+7.95}_{-5.49}$        & 4.34$^{+2.57}_{-0.6}$   & 10.99$^{+0.09}_{-0.06}$ & BPT  \\
					GMBCG J143.35190+06.97592 & 143.351897556082 & 6.9759245772116    & 0.19$\pm$0.02 & 0.0       & 8  & 20.08$\pm$0.12 & 18.74$\pm$0.02 & 17.75$\pm$0.01 & 17.25$\pm$0.01 & 16.92$\pm$0.03 & 16.68$\pm$0.01 & 16.91$\pm$0.02 & 15.12$\pm$0.05  & 14.08$\pm$0.12 & 11.68$^{+0.24}_{-0.42}$ & 48.26$^{+36.23}_{-30.03}$     & 6.47$^{+2.69}_{-2.52}$  & 10.99$^{+0.13}_{-0.14}$ & NONE \\
					GMBCG J145.91519+08.93535 & 145.91518899381  & 8.93534998424573   & 0.1$\pm$0.03  & 0.103973  & 11 & 18.06$\pm$0.03 & 16.87$\pm$0.01 & 16.23$\pm$0    & 15.86$\pm$0    & 15.59$\pm$0.01 & 15.67$\pm$0    & 16.12$\pm$0.01 & 14.17$\pm$0.02  & 13.91$\pm$0.11 & 10.93$^{+0.59}_{-0.06}$ & 8.52$^{+24.81}_{-1.12}$       & 4.02$^{+0.65}_{-1.07}$  & 10.51$^{+0.09}_{-0.05}$ & NONE \\
					GMBCG J145.99552+04.28192 & 145.995520564657 & 4.28191745024208   & 0.29$\pm$0.09 & 0.0       & 10 & 19.91$\pm$0.2  & 19.13$\pm$0.02 & 18.01$\pm$0.02 & 17.57$\pm$0.01 & 17.25$\pm$0.05 & 16.65$\pm$0.01 & 16.81$\pm$0.02 & 15.62$\pm$0.08  & 14.54$\pm$0.19 & 11.88$^{+0.27}_{-0.37}$ & 75.11$^{+66.11}_{-42.97}$     & 5.6$^{+2.09}_{-2.32}$   & 11.31$^{+0.09}_{-0.11}$ & NONE \\
					GMBCG J147.22126+27.98813 & 147.221264746376 & 27.9881295516909   & 0.41$\pm$0.09 & 0.0       & 16 & 21.68$\pm$0.29 & 20.95$\pm$0.07 & 19.69$\pm$0.03 & 19.08$\pm$0.03 & 18.74$\pm$0.07 & 17.82$\pm$0.02 & 17.33$\pm$0.03 & 15.71$\pm$0.08  & 14.49$\pm$0.17 & 12.21$^{+0.26}_{-0.34}$ & 163.08$^{+136.56}_{-87.87}$   & 6.06$^{+1.85}_{-2.42}$  & 11.2$^{+0.09}_{-0.11}$  & WISE \\
					GMBCG J148.38032+22.74570 & 148.380320659692 & 22.7456989645336   & 0.17$\pm$0.02 & 0.20488   & 15 & 19.52$\pm$0.08 & 18.17$\pm$0.01 & 17.25$\pm$0.01 & 16.81$\pm$0.01 & 16.5$\pm$0.02  & 16.47$\pm$0.01 & 16.68$\pm$0.02 & 15.02$\pm$0.05  & 14.56$\pm$0.18 & 11.91$^{+0.08}_{-0.49}$ & 80.91$^{+17.02}_{-54.94}$     & 4.98$^{+2.85}_{-2.68}$  & 11.14$^{+0.09}_{-0.14}$ & BPT  \\
					GMBCG J149.74544+21.13741 & 149.745443703615 & 21.1374109487258   & 0.29$\pm$0.03 & 0.0       & 17 & 19.55$\pm$0.07 & 18.95$\pm$0.01 & 18.21$\pm$0.01 & 17.88$\pm$0.01 & 17.51$\pm$0.04 & 17.29$\pm$0.01 & 17.42$\pm$0.03 & 15.65$\pm$0.07  & 13.79$\pm$0.08 & 11.66$^{+0.28}_{-0.15}$ & 45.98$^{+41.68}_{-13.77}$     & 1.07$^{+1.61}_{-0.21}$  & 10.68$^{+0.09}_{-0.07}$ & NONE \\
					GMBCG J151.66740+21.67074 & 151.66739864     & 21.67074124        & 0.19$\pm$0.01 & 0.189074  & 19 & 19.62$\pm$0.16 & 17.46$\pm$0.01 & 16.21$\pm$0    & 15.7$\pm$0     & 15.34$\pm$0.01 & 15.19$\pm$0    & 15.43$\pm$0.01 & 15.82$\pm$0.16  & 14.06$\pm$0.18 & 11.22$^{+0.45}_{-0.18}$ & 16.63$^{+30.61}_{-5.69}$      & 5.01$^{+2.37}_{-0.69}$  & 11.62$^{+0.06}_{-0.06}$ & BPT  \\
					GMBCG J151.86216+29.50568 & 151.86215965     & 29.50568309        & 0.15$\pm$0.03 & 0.116862  & 15 & 19.18$\pm$0.07 & 17.55$\pm$0.01 & 16.61$\pm$0    & 16.12$\pm$0    & 15.85$\pm$0.01 & 15.79$\pm$0.01 & 16.11$\pm$0.01 & 15.73$\pm$0.12  & 13.92$\pm$0.15 & 10.78$^{+0.43}_{-0.19}$ & 6.05$^{+10.13}_{-2.17}$       & 4.18$^{+4.03}_{-0.45}$  & 10.68$^{+0.43}_{-0.05}$ & NONE \\
					GMBCG J151.96927+27.50322 & 151.9692698      & 27.50322434        & 0.15$\pm$0.02 & 0.148337  & 9  & 18.31$\pm$0.03 & 16.97$\pm$0    & 16.12$\pm$0    & 15.71$\pm$0    & 15.42$\pm$0.01 & 15.39$\pm$0    & 15.74$\pm$0.01 & 14.27$\pm$0.02  & 14.1$\pm$0.12  & 11.79$^{+0.07}_{-0.48}$ & 61.01$^{+10.79}_{-40.71}$     & 3.49$^{+4.02}_{-0.48}$  & 11.19$^{+0.05}_{-0.05}$ & NONE \\
					GMBCG J152.05418+12.99533 & 152.05418172     & 12.99532823        & 0.47$\pm$0.02 & 0.0       & 8  & 22.52$\pm$0.98 & 21.04$\pm$0.09 & 19.44$\pm$0.03 & 18.59$\pm$0.03 & 18.19$\pm$0.08 & 17$\pm$0.01    & 17.46$\pm$0.04 & 17.17$\pm$0.34  & 14.4$\pm$0.15  & 11.9$^{+0.4}_{-0.16}$   & 79.51$^{+120.16}_{-24.87}$    & 6.15$^{+1.58}_{-2.52}$  & 11.65$^{+0.08}_{-0.14}$ & NONE \\
					GMBCG J152.23915+56.43633 & 152.23914718     & 56.43632872        & 0.27$\pm$0.07 & 0.0       & 21 & 21.24$\pm$0.21 & 20.09$\pm$0.03 & 18.86$\pm$0.02 & 18.29$\pm$0.02 & 17.87$\pm$0.03 & 17.46$\pm$0.01 & 17.72$\pm$0.03 & 16.53$\pm$0.13  & 14.91$\pm$0.2  & 11.31$^{+0.44}_{-0.18}$ & 20.27$^{+35.15}_{-6.95}$      & 7.08$^{+1.87}_{-2.05}$  & 11.05$^{+0.1}_{-0.1}$   & NONE \\
					GMBCG J153.24029+17.50484 & 153.24028518     & 17.50484024        & 0.12$\pm$0.02 & 0.115981  & 9  & 18.55$\pm$0.06 & 16.77$\pm$0.01 & 15.81$\pm$0    & 15.37$\pm$0    & 15.03$\pm$0.01 & 15.16$\pm$0    & 15.66$\pm$0.01 & 14.84$\pm$0.05  & 14.39$\pm$0.19 & 11.05$^{+0.28}_{-0.38}$ & 11.35$^{+10.09}_{-6.56}$      & 7.64$^{+1.15}_{-1.05}$  & 11.11$^{+0.05}_{-0.05}$ & BPT  \\
					GMBCG J153.57351+47.87017 & 153.57351363     & 47.87017286        & 0.29$\pm$0.04 & 0.0       & 11 & 19.03$\pm$0.04 & 18.38$\pm$0.01 & 17.5$\pm$0.01  & 17.17$\pm$0.01 & 16.85$\pm$0.02 & 16.75$\pm$0.01 & 16.87$\pm$0.02 & 15.1$\pm$0.04   & 14.86$\pm$0.19 & 12.26$^{+0.05}_{-0.24}$ & 183.23$^{+21.7}_{-78.06}$     & 4.32$^{+2.19}_{-2.08}$  & 11.15$^{+0.12}_{-0.12}$ & NONE \\
					GMBCG J154.11671+46.51435 & 154.11671456     & 46.51435331        & 0.15$\pm$0.04 & 0.0       & 8  & 17.56$\pm$0.01 & 17.04$\pm$0    & 16.5$\pm$0     & 16.09$\pm$0    & 15.91$\pm$0.01 & 15.43$\pm$0    & 15.35$\pm$0    & 14.42$\pm$0.03  & 14.02$\pm$0.11 & 11.82$^{+0.06}_{-0.27}$ & 66.74$^{+10.26}_{-30.75}$     & 8                       & 11.02                   & NONE \\
					GMBCG J154.28803+39.96277 & 154.2880316      & 39.96276587        & 0.24$\pm$0.03 & 0.0       & 12 & 21.16$\pm$0.16 & 19.87$\pm$0.02 & 18.68$\pm$0.01 & 18.21$\pm$0.01 & 17.87$\pm$0.02 & 17.57$\pm$0.02 & 17.82$\pm$0.04 & 16.72$\pm$0.17  & 14.77$\pm$0.18 & 11.13$^{+0.44}_{-0.18}$ & 13.4$^{+23.37}_{-4.53}$       & 7.14$^{+2.02}_{-3.22}$  & 10.93$^{+0.14}_{-0.14}$ & NONE \\
				\end{tabular}}
				\begin{tablenotes}
					\item[\textdagger] As quoted from the GMBCG catalog
					\item	Note: All photometry is in AB magnitude.
				\end{tablenotes}
			\end{threeparttable}
		\end{sidewaystable*}
\end{document}